%% file: bru3Dr1.tex
\documentclass[a4paper,11pt]{article}
\usepackage[top=25mm,left=22mm]{geometry}
\usepackage{amsmath,amssymb,amsthm,amsfonts}
\usepackage{lineno,url}\usepackage{float}
\floatstyle{plaintop}\restylefloat{table}
\usepackage[tableposition=top]{caption}
\usepackage{graphicx}%,epsfig}
\usepackage{multirow}\usepackage{pstricks}\usepackage{fancyvrb}
\usepackage{listings,paralist, longtable}
\newtheorem{theorem}{Theorem}[section]
\theoremstyle{definition}
\newtheorem{remark}[theorem]{Remark}
\input{hudefmin.tex}
\def\huc#1{#1}%\def\huc#1{{\red #1}}
\def\pdep{{\tt pde2path}}
\newlength{\tew}

\def\huga#1{\begin{gather} #1 \end{gather}}

\providecommand{\ali}[1]{\begin{align}#1\end{align}}
\def\bci{\begin{compactitem}}\def\eci{\end{compactitem}}

\def\ig{\includegraphics}

\def\medskip{}\def\bigskip{}

\def\ass{{\tt ampsys}}
\def\Ab{A^{\text{bcc}}}\def\At{A^{\text{tube}}}

\def\hual#1{\begin{align} #1 \end{align}}

\def\eex{\hfill\mbox{$\rfloor$}}
\def\vx{\vec x}\def\vk{\vec k}
\newcommand{\btab}[2]{\begin{tabular}{#1}#2\end{tabular}}

\usepackage{hyperref}
\begin{document}
\title{Snaking branches of planar BCC fronts in the 3D Brusselator}
\author{Hannes Uecker\footnote{Institut f\"ur Mathematik, Universit\"at Oldenburg, D-26128 Oldenburg, Germany; hannes.uecker@uol.de}, \ \ 
Daniel Wetzel\footnote{danieldwetzel@gmail.com}}
\maketitle
\begin{abstract}
We present results 
of the application of the numerical continuation and bifurcation package 
\pdep\ to the 3D Brusselator model, focusing on snaking branches of 
planar fronts between body centered cubes (BCCs) and the 
spatial homogeneous solution, and on planar fronts between BCCs 
and tubes (also called prisms). 
These solutions also yield approximations of localized BCCs, and 
of BCCs embedded in a background of tubes (or vice versa). Additionally, 
we compute some moving fronts between lamellas and tubes. 
To give some theoretical background, and to aid 
the numerics for the full system, we use the Maxwell points for the 
cubic amplitude system over the BCC lattice. 
\end{abstract}
\noi
\section{Introduction}\label{isec}
Turing patterns \cite{tu52} are stationary solutions of Reaction--Diffusion 
PDE systems that bifurcate from a homogeneous steady state which becomes unstable with respect to perturbations with a 
critical wave number $k_c\ne 0$. 
If the bifurcation is subcritical and the bifurcating 
branch stabilizes in a fold, then this gives bistability between the 
homogeneous state and the larger amplitude patterns in the subcritical 
regime, and this often yields the existence of localized patches of 
periodic patterns. These localized patterns exist in 
extended regions in parameter space \cite{pomeau}, and come in 'snaking' 
branches which move back and forth in parameter space. 
This mechanism is well studied in the one--dimensional and two--dimensional cases (1D and 2D, respectively), see, e.g., 
\cite{burke,bukno2007,BKLS09,strsnake,KUW19} for mainly numerical 
results, and \cite{chapk09,dean11,KC13,deWitt19} for analysis  
using the Ginzburg-Landau formalism and beyond all order asymptotics.

1D patterns extended homogeneously into a second and third direction are also solutions over 2D and 3D domains and are then referred to as {\em stripes} and {\em lamellas}, 
respectively. Typical genuine 2D patterns are {\em squares} and {\em hexagons}, and extended homogeneously in a third direction these 
yield (square and hexagon, respectively) {\em tubes}, while the simplest genuine 
3D periodic patterns are {\em cubes} (or balls). Numerically, 
Turing patterns in 3D 
have so far mostly been studied 
by direct numerical simulation (DNS, aka numerical time integration)  
\cite{WBD97, SYUO07}. %, but we are not aware of systematic numerical branch following and bifurcation studies in 3D. 
Additionally, some experimental results (and associated DNS for the Lengyel Epstein model) are reported in \cite{EV11}.  
See also \cite{alberetal05,glimm08} for further results and discussion. 

In \cite{uwsnak14} we numerically study planar fronts between stripes and hexagons in the 2D Schnakenberg model, using the 
package \pdep\ \cite{p2pure,p2phome}. Such fronts (or heteroclinic connections) can be naturally extended to localized patterns (or homoclinic cycles) by 
gluing together fronts and backs (i.e., considering heteroclinic cycles). 
See also, e.g., \cite{w16, w18} for various further results on localized 2D patterns in different reaction-diffusion systems, and the Swift--Hohenberg equation as another prototype pattern forming 
system. Moreover, \cite{w16} also contains a brief 
outlook on branches of 3D patterns, and some results on 3D patterns including localized patterns 
are also explained in \cite{pftuta}, with detailed explanations on 
the background and usage of \pdep. 

In a similar fashion as in \cite{uwsnak14} in 2D, here we study 3D 
planar fronts between cubes and 
the homogeneous steady state, and between cubes and tubes, 
in the Brusselator model \cite{brus}
\begin{equation}\label{bruss1}
\begin{aligned}
\pa_t{u_1}&=-(B+1)u_1+u_1^2u_2+A+D_1\Delta u_1,\\
\pa_t{u_2}&=Bu_1-u_1^2u_2+D_2 \Delta u_2. 
\end{aligned}
\end{equation}
The chemical concentrations $u_1=u_1(\vx,t)$ and $u_2=u_2(\vx,t)$, 
with spatial coordinate $\vx=(x_1,x_2,x_3)\in\Om\subset\R^3$ and time $t\ge 0$,  
correspond to an activator and inhibitor, respectively, 
$D_1$ and $D_2$ are their diffusivities, $A$ and $B$ are parameters, 
and $\Delta$ is the Laplacian. \huc{We let $u=(u_1,u_2)$, and 
instead of the coordinates $(x_1,x_2,x_3)$ of $\vx$ 
we also write $(x,y,z)$.} 
Moreover, \reff{bruss1} has to be 
complemented by suitable boundary conditions (BCs) on $\pa\Om$, and we will 
restrict to homogeneous Neumann BCs, i.e., 
\huga{\label{bc1}
\pa_\nu u_1=\pa_\nu u_2=0\text{ on }\pa\Om,
}
where $\pa_\nu$ denotes the outer normal derivative. For the initial value 
problem we also need to prescribe initial conditions $u|_{t=0}=u_0$. 

Homogeneous steady states of \reff{bruss1} are given 
by $u_1^*=A$ and $u_2^*=B/A$. 
We fix the parameters 
\huga{\label{pchoice}
A=2, \quad D_1=1, \quad D_2=(A/R)^2, 
} 
where $R$ is used as a convenient unfolding parameter, see below, 
and consider $B$ as the primary bifurcation parameter. 
The steady state $u^*=(u_1^*,u_2^*)$ is stable 
for 
\huga{\label{bcdef}
B<B_c=(1+R)^2,
} 
where a Turing bifurcation occurs with critical wave number $k_c=\sqrt{R}$. 
Our focus will be on solution branches corresponding to a so called body centered cube (BCC) lattice. Close to bifurcation, these may be described 
by a system of equations for six amplitudes $A=(A_1,\ldots,A_6)\in\C^6$, 
see \reff{redbcc} in \S\ref{afsec}. This amplitude system 
has a variety of steady solutions $A^*$, which  in the original system \reff{bruss1} correspond to, e.g., 
lamellas, tubes, and cubes, henceforth called BCCs. These solution branches of the amplitude equations have been classified and discussed in detail in \cite{CKnob97, CKnob99}, and the stability of the associated solutions of the original system close to onset 
has been studied in \cite{CKnob01}. See also \cite{GoS2002, Hoyle} for 
textbook expositions of the underlying and very important symmetry perspective. 
Additionally, the (ODE) amplitude system can be formally extended to 
a (1D PDE) modulation equation system by assuming a slow dependence of the 
amplitudes on one spatial coordinate. The steady modulation equations 
have a spatially conserved quantity (the potential energy) which thus 
defines Maxwell points for heteroclinics between different fixed 
points $A^*$. These results can then be used to identify parameter regimes 
for the search of snaking branches of steady 
fronts between BCCs and $u^*$, and between BCCs and tubes. 
For this, to keep the numerics inexpensive we choose small $\Om$, 
e.g.~boxes $\Om=(-l,l)^2\times (-l_z,l_z)$, where $l=\pi/k_c$  
$l_z=4l$. Additionally, we briefly illustrate that near to but outside 
the snaking region the dynamics of close by solutions show a stick--slip 
motion, and give examples of other moving fronts, for instance between 
lamellas and cubes.

\brem{\rm Our results are not specific to the Brusselator model \reff{bruss1}, 
but can be expected for all 3D pattern forming systems with a subcritical 
Turing bifurcation, or, more generally, systems with a bistability 
of patterns and the homogeneous solutions, or a bistability 
of different patterns. 
Similar results are provided for the (quadratic--cubic) 3D Swift--Hohenberg  equation in 
\cite[\S3]{pftuta}. There we also give detailed explanations on a number of 
issues that arise for numerical branch continuation and bifurcation 
in 3D pattern forming systems, including:
\bci
\item The algorithm for branch switching at branch points of higher 
multiplicity, which naturally arise in 3D due to symmetries, see also 
\cite{pftuta}. 
\item Tricks, including remarks on the choice of meshes, to avoid uncontrolled 'branch jumping', which is a major issue 
in particular in 3D due to the multitude of different branches close to 
each other. 
\eci 
Here we focus on the Brusselator model \reff{bruss1} as a standard reaction--diffusion model. 
\eex}\erem 

\noi
   {\bf Acknowledgment.}  The work of DW was supported by the DFG under Grant No.~264671738.

\section{The amplitude formalism}\label{afsec}
We briefly review the BCC amplitude equations 
for \reff{bruss1} close to the primary bifurcation from $u^*$, focusing 
on the bifurcating branches pertaining to Neumann BCs over cuboids. 
Amplitude equations for 3D pattern forming systems are derived 
and discussed in \cite{CKnob97, CKnob99} for three lattices with cubic symmetry, namely the simple cubic, the face-centered cubic, and the body-centered cubic (BCC). 
\subsection{Derivation of the amplitude system}
For the BCC lattice, the critical wave vectors are given by
\huga{\label{wav}
\barr{lll}
\vk^{(1)}=-\vk^{(7)}=\frac{k_c}{\sqrt{2}}(1,1,0),& \vk^{(2)}=-\vk^{(8)}=\frac{k_c}{\sqrt{2}}(0,1,1), 
&\vk^{(3)}=-\vk^{(9)}=\frac{k_c}{\sqrt{2}}(1,0,1),\\
\vk^{(4)}=-\vk^{(10)}=\frac{k_c}{\sqrt{2}} (1,-1,0), &\vk^{(5)}=-\vk^{(11)}=\frac{k_c}{\sqrt{2}} (0,1,-1), &\vk^{(6)}=-\vk^{(12)}=\frac{k_c}{\sqrt{2}}(-1,0,1), 
\earr
}
where $k_c=\sqrt{R}$ is the critical wave number. 
Setting $u=u^*+w$, $w=w(\vx,t)\in\R^2$, yields 
\huga{\label{weq}
\pa_t w=Lw+N(w),
} 
where $L$ is the linearization around $u^*$ and $N(w)$ denotes the 
nonlinear terms. We make the ansatz 
\ali{\label{ansatz}
w(\vx,t)=\sum_{j=1}^{6}A_j(t)e^{\text{i}\vk^{(j)} \cdot \vx}\Phi+\text{c.c.}+\text{h.o.t}, 
}
where $A_j\in\C$, and $\Phi=(\Phi_1,\Phi_2)\in\C^2$ is the critical eigenvector, 
independent 
of $j$ due to the rotational invariance of the Laplacian, and normalized to $\Phi_1=1$. In \reff{ansatz}, c.c.~means the 
complex conjugate of the preceding terms, 
and h.o.t.~denotes higher order terms, which turn out to 
be nonlinear terms in the $A_j$. Plugging \reff{ansatz} into \reff{weq}, 
sorting wrt.~to the modes $\er^{\ri \vk^{(j)}\cdot \vx}$, 
first solving for uncritical modes at, e.g., $e_0$, $e_{2k_1}$ and so on,  
we obtain the amplitude equations 
\ali{\label{redbcc}
\dot{A_i}=f_i(A_1,A_2,A_3,A_4,A_5,A_6), \quad i=1,\ldots,6.  
}
Their general form, dictated by symmetry \cite{CKnob97}, is 
\huga{\begin{aligned}
f_1=&\lam A_1+q(A_2\ov{A}_6+A_3A_5)+
 c_{31}|A_1|^2 A_1+c_{32}(|A_2|^2+|A_3|^2+|A_5|^2+|A_6|^2)A_1\\
 &+c_{33}|A_4|^2A_1+c_{34}(A_2A_4A_5+A_3\ov{A}_4\ov{A}_6), \\
f_2=&\lam A_2+q(A_1A_6+A_3\ov{A}_4)+
 c_{31}|A_2|^2 A_2+c_{32}(|A_1|^2+|A_3|^2+|A_4|^2+|A_6|^2)A_2\\
 &+c_{33}|A_5|^2A_2+c_{34}(A_1\ov{A}_4\ov{A}_5+A_3A_5A_6), 
\end{aligned}
 \label{bccgen2}
}
and the remaining $f_j, j=3,\ldots,6$ (we shall not need them explicitly) 
also follow from symmetry. The BCC lattice supports three-wave 
interactions, e.g., $k^{(1)}=k^{(2)}-k^{(6)}=k^{(3)}+k^{(5)}$, which explains the occurrence of the quadratic terms $q(A_2\ov A_6+A_3A_5)$ in $f_1$. For the coefficients 
$\lam$, $q$ and $c_{31}$ we have some analytic formulas which follow 
from, e.g., \cite{VWDB92}, see also \cite{CKnob99}, namely 
\huga{\label{lamcoeff}
\lam= \del(B-B_c)\text{ with }\del=\frac{A^2}{(A^2-R^2)(R+1)},\quad 
q=\frac{2A(1-R)}{A^2-R^2},\quad c_{31}=\frac{8-38R-5R^2+8R^3}{9R(A^2-R^2)}. 
}
Similar formulas can be derived for $c_{32},c_{33}$ and $c_{34}$, but we refrain 
from doing so, and instead will use numerical values computed by 
the \pdep\ tool \ass\ \cite{ampsys}, which is designed to do such computations 
with minimal user input. %
\footnote{In \cite{ampsys} we apply our tool to 
a variety of models and wave vector lattices, and cases with known 
coefficients such as \reff{lamcoeff} are useful for checks 
of the implementation. Conversely, using {\tt ampsys} on known 
cases is helpful to make sure that scalings of amplitudes are 
taken care of correctly. For instance, 
the formulas for $q$, $c_{31}$ in \cite{CKnob99} are 
different from \reff{lamcoeff} because they are based on a 
normalization of the critical eigenvector $\Phi=(\Phi_1,\Phi_2)$ 
which is different 
from our normalization with $\Phi_1=1$.} 

Quadratic and cubic terms are considered to be of 
the same order for the derivation of \reff{redbcc}, and this formally 
requires $q$ to be small, which means $R\approx 1$, cf.~Remark \ref{hexrem}. 
Moreover, as in \cite{CKnob99}, 
\huga{\label{c3jlim}
\text{$\lim_{R\to 1}c_{32}/c_{31}=\lim_{R\to 1}c_{33}/c_{31}=
\lim_{R\to 1}c_{34}/c_{31}=2$,}
}
and $\lim_{R\to 1}c_{31}= -1$ for the choice $A=2$, 
which we fix in the numerics. 
However, we shall be interested in $1-R=\CO(1)$, and 
the deviations of $c_{32}, c_{33}$ and $c_{34}$ from 
$2c_{31}$ turn out to be significant in this case. 

The bifurcation diagrams 
for \reff{redbcc} with $B$ close to $B_c$, 
have been discussed in detail in \cite{CKnob97, CKnob99}. 
Here we restrict to those branches that fulfill Neumann BC on cuboid 
domains of the form $\Om=(-l_x,l_x)\times(-l_y,l_y)\times(-l_z,l_z)$ with 
$l_x=n_1 l, l_y=n_2 l, l_z=n_3 l$, $n_j\in\N/2$ and $l=\sqrt{2}\pi/k_c$. 
By \reff{wav} and \reff{ansatz}, this restricts (modulo phase-shifts, 
i.e., spatial translations by $nl$ for some $n\in\N/2$) 
the admissible solutions to 
$$
(A_1,A_2,A_3,A_4,A_5,A_6)=(A_1,A_2,A_2,A_1,A_2,A_2), 
$$
where $A_1,A_2\in\R$ 
fulfill
\huga{\label{reda}
\barr{rl}
\dot A_1&=\lam A_1+2qA_2^2+(\al A_1^2+2\beta A_2^2)A_1,\\
\dot A_2&=\lam A_2+2qA_1A_2+(\beta A_1^2+\ga A_2^2)A_2, 
\earr
}
with the effective coefficients 
\huga{\label{c3j}
\al=c_{31}+c_{33}, \quad \beta=2c_{32}+c_{34}, \quad 
\ga=c_{31}+2c_{32}+c_{33}+c_{34}.
}
In Table \ref{lctab} we list these coefficients 
(together with further data explained below) 
for some chosen values of $R$, 
for which we shall also run numerics on the full system \reff{bruss1}. 

\begin{table}[ht]
\caption{Landau coefficients and other data (see Remark \ref{hexrem}) 
for \reff{reda}, $A=2$.\label{lctab}}
\bce\vs{-4mm}
{\small 
\begin{tabular}{c|ccccc|ccc}
$R$&$\del$&$q$&$\al$&$\beta$&$\ga$&$\eps=B_c-B_f$&$B_M$&$\tilde{B}_M$\\
\hline
1&2/3&0&-3&-6&-9&0&NA&NA\\
0.75&0.665&0.29&-1.2&-5.95&-7.15&0.01&3.054&3.21\\
0.52&0.532&0.515&1.88&-5.1&-3.2&0.05&2.266&NA\\
0.4&0.476&0.625&4.7&-3.78&0.93&0.18&1.82&NA\\
\end{tabular}
}
\ece

\vs{-8mm}
\end{table}
\brem\label{srrem}{\rm 
We decrease $R$ rather far from $R=1$ to $R=0.4$, and, moreover, 
will use \reff{reda} for $B-B_c=\CO(1)$. Each operation alone  
makes the applicability of \reff{reda} quite questionable. However, 
some of the interesting results will occur in the strongly subcritical 
regime $R=0.4$, and applying \reff{reda} with care we get good 
predictions for these, while other effects cannot be captured. 
See, e.g., the discussion of Fig.~\ref{hf4b} below, and 
\cite{BMS09} for a related general discussion about the use 
of amplitude equations for subcritical bifurcations. \eex}\erem

\subsection{Steady solutions}
\begin{figure}[ht]
\bce 
{\small
\begin{tabular}{ll}
(a)&(b) tube\hs{20mm}(c) hot BCC\\
%\hs{-5mm}\ig[width=0.22\tew,height=50mm]{hp2/gbd75e}&
\hs{-5mm}\ig[width=0.24\tew]{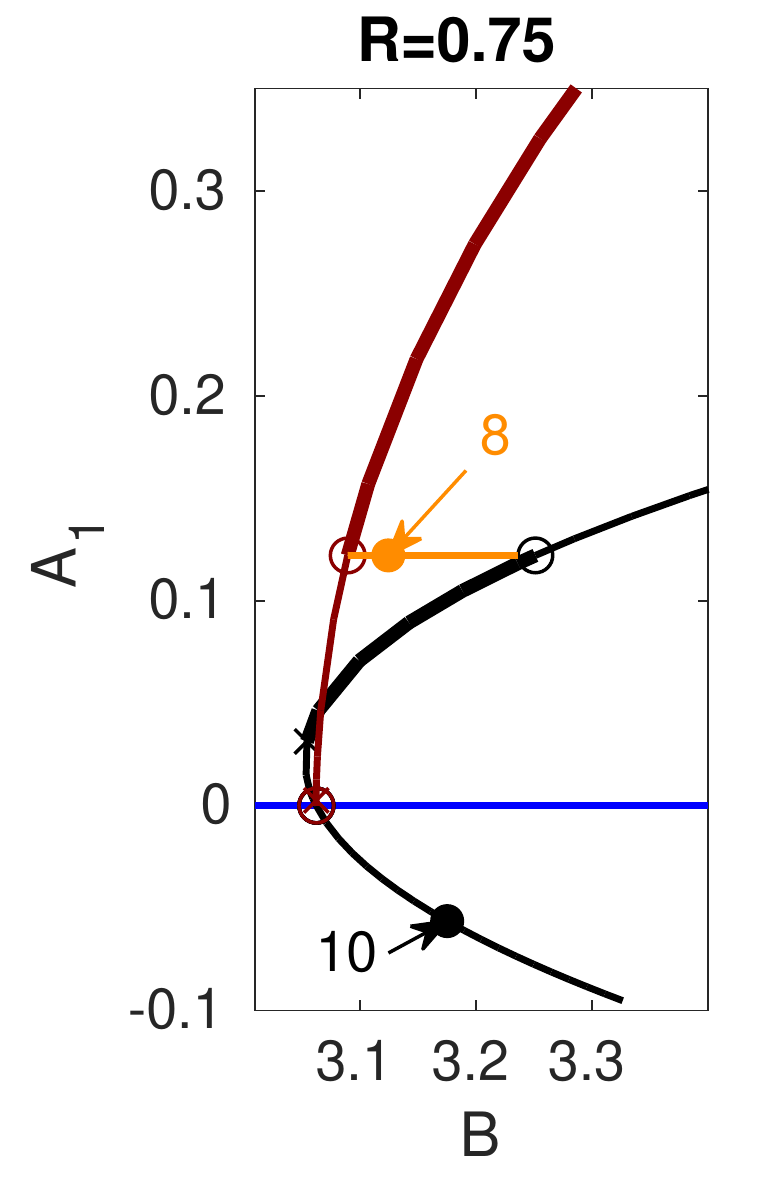}&
\raisebox{24mm}{\btab{ll}{\ig[width=0.18\tew]{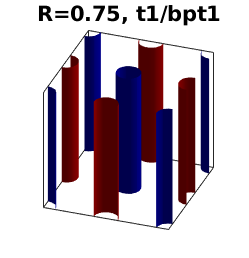}
&\ig[width=0.18\tew]{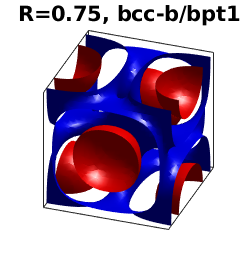}\\
(d) mixed&(e) cold BCC\\
\ig[width=0.18\tew]{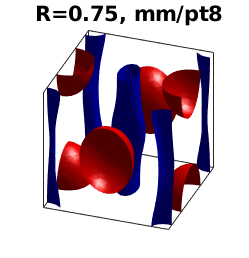}&\ig[width=0.18\tew]{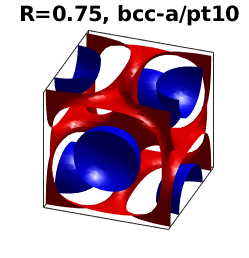}}}
\end{tabular}
}
\ece 

\vs{-5mm}
\caption{{\small (a) Solution branches   of the amplitude system \reff{reda} for \reff{bruss1} with $A=2$ and $R=0.75$, as functions of $B$, 
using $\lam=\del(B-B_c)$; trivial branch (blue), tubes (red), 
BCCs (black), and mixed modes (orange); stable parts in thicker lines. 
(b)-(d): Sample solutions as reconstructed via \reff{ansatz} (with a phase-shift for better illustration and h.o.t.~truncated) from the labeled points in (a), i.e., from the mixed mode 
branch {\tt m1/pt20}, from the branch points connected by it, and from 
the cold BCCs.  
Isosurfaces $w_1=-c$ (blue) and $w_1=c$ (red) on $\Om=(-l,l)^3, 
l=\sqrt{2}\pi/k_c$. $c=0.4$ in (b)--(d),  $c=0.1$ in (e). 
\label{hf4}}}
\end{figure}

The solution $A_1=A_2=\Ab_{\pm}$ of \reff{reda} with 
%\huga{
$\Ab_\pm=-\frac{q}{\al+2\beta}\pm \sqrt{\frac{q^2}{(\al+2\beta)^2}-
\frac{\lambda}{\al+2\beta}}$
%} 
yields BCCs for \reff{bruss1} 
\def\wbcc{w_{{\rm BCC}}}\def\wtube{w_{{\rm tube}}}
in the form 
\huga{\label{bccf1} 
\begin{aligned}
\wbcc=2\Ab\biggl[&\cos(\kappa(x+y))+\cos(\kappa(y+z)+\cos(\kappa(x+z))\\
&+\cos(\kappa(x-y))+\cos(\kappa(y-z))+\cos(\kappa(-x+z))\biggr]\Phi+\text{h.o.t.}\end{aligned}, 
}
and naturally phase shifts in $A_j$ correspond to translations in $x,y,z$, 
which we shall not distinguish from \reff{bccf1}.   
The BCC branch bifurcates supercritically if $q=0$ ($R=1$) 
and transcritically if $q\neq 0$ ($R<1$), and has a fold in 
\huga{\label{bccf}
B_f=B_c+\frac{q^2}{(\al+2\beta)\del}. 
}  
The system \reff{redbcc} is equivariant 
under $q\mapsto -q$ and $A_i\mapsto -A_i$, $i=1,\ldots,6$, and this is naturally 
inherited by \reff{reda}. 
 Depending on the sign of $q$, one direction of the BCCs has maxima 
of $u_1$ in the centers of the balls, and we call these {\em 'hot'} balls, 
while in the other direction we have {\em 'cold'} balls, see Fig.~\ref{hf4}(d,e).  This classification 
is analogous to 'spots' and 'gaps' in the 2D case. 

The other primary solution branch of \reff{reda} (i.e., of \reff{redbcc} and compatible with Neumann BCs) yields 'tubes' 
(called 'squares' in \cite{CKnob97} and in much of the literature), i.e.,
%\huga{
$A_1=\At_\pm:=\pm\sqrt{-\frac{\lam}{\al}}, A_2=0$,
%}
and these bifurcate in pitchforks. The associated solutions 
$w$ of \reff{weq} are spatially homogeneous in $z$ direction, namely 
\huga{\label{tf1} 
\wtube=2\At[ \cos(\kappa(x+y))+\cos(\kappa(x-y))]\Phi+\text{h.o.t.},  
}
see Fig.~\ref{hf4}(b). From $\al$ in Table \ref{lctab} we readily 
see that the bifurcation of the tubes changes from super-- to subcritical for $R$ 
decreasing from $0.75$ to $0.52$. 

For simplicity, we shall also denote the vectors $\vec{A}=(A_1,A_2)$ 
similarly, i.e., 
\def\Abv{\vec{A}^{\text{bcc}}}\def\Atv{\vec{A}^{\text{tube}}}
\huga{\label{asym}
\Abv_\pm:=(\Ab_\pm,\Ab_\pm)\text{ and }
\Atv_\pm:=(\At_\pm,0). }
For $R\ne 1$, both of these branches (families of branches, via symmetries), 
$\wbcc$ and $\wtube$, are unstable 
close to bifurcation. However, 
on the level of the amplitude equations, the BCCs stabilize after the 
fold, and the tubes at an $O(1)$ distance from onset, while the stable BCC branch again destabilizes at $O(1)$ distance, and there is an 
unstable {\em mixed mode} branch connecting the BCCs and the tubes. 

In Fig.~\ref{hf4}(a) we illustrate the branching 
behavior for $R=0.75$ of three nontrivial branches of \reff{reda}. 
Additional to the BCCs (black) and tubes (red) there is the mixed mode branch 
(orange) connecting tubes and BCCs between the points where these gain/loose 
stability, i.e., in their bistable range. 
In particular, in the amplitude equations we get two 
bistabilities: (a) bistability of $A=0$ and the $\Ab_+$ part below 
onset, and (b) bistability of $\Ab_+$ and $\At$ at $O(1)$ distance 
above onset. This does in 
general not mean that the associated solutions inherit these 
in the full system \cite{CKnob01}. However, this turns out roughly to be the case
over sufficiently small domains (which may be significantly extended in $z$). 
This motivates our main aim, i.e., to find snaking branches of 
localized BCCs, or more precisely,  of fronts (a) 
between BCCs and $u=u^*$, and (b) between BCCs and tubes. 

\brem\label{hexrem}{\rm 
(a) The reduced amplitude equations \reff{reda} have exactly the same structure 
as the amplitude equations for the three modes $(A_1,A_2,A_3)$ in the 
2D case on a planar hexagonal lattice when restricted to the subspace 
$A_2=A_3$, see, e.g., \cite[\S3.1]{uwsnak14}. 
The 3D tubes and BCCs thus correspond to 2D stripes and hexagons, 
respectively, and the stabilities within the amplitude system are also 
equivalent. However, the stability and bifurcation structures of the associated 
solutions in the original 2D vs 3D systems will in general be rather 
different. 

(b) The 7th column of Table \ref{lctab} indicates how the 
'subcriticality' $\eps:=B_c-B_f=q^2/((\al+2\beta)\del)$, cf.~\reff{bccf}, 
of the BCC branch increases with decreasing $R$. 
In, e.g., \cite{chapk09,dean11} it is explained 
(for Swift--Hohenberg models) that 
the snaking width of branches of fronts connecting a subcritical pattern 
and $0$ is exponentially small in this subcriticality $\eps$, 
which for \reff{bruss1} means that we expect that $|B_l-B_r|\sim c_1\eps^{-1}
\er^{-c_2/\eps}$ where $c_1,c_2$ are constants and $B_l$ and $B_r$ denote 
the left and right ends of the snaking range, respectively. 
Relatedly, the steepness of the fronts scales as $\eps$, and hence the 
required 
domain length as $1/\eps$. 
See also \cite{uwsnak14} for numerical illustrations of this phenomenon, 
and \cite{deWitt19} for further references and a transfer of the results from 
\cite{chapk09,dean11} to reaction diffusion systems. 
Thus, if we assume that snaking branches between BCCs and $u=u^*$ exist, then Table \ref{lctab} also indicates that finding these should be more robust and less expensive at smaller $R$. The 8th column gives (for $R<1$) the approximate 
Maxwell point between BCCs and $0$ (see \S\ref{maxsec}), 
and the 9th column the one between 
BCCs and tubes (which for the used parameters only exists in the second row). 
}\eex\erem 

\subsection{Maxwell points in the amplitude system}\label{maxsec}
\begin{figure}[ht]
\bce 
{\small
\begin{tabular}{llll}
(a) $R=1$& (b) $R=0.75$&(c) $R=0.52$&(d) $R=0.4$\\
\hs{-0mm}\ig[width=0.2\tew,height=50mm]{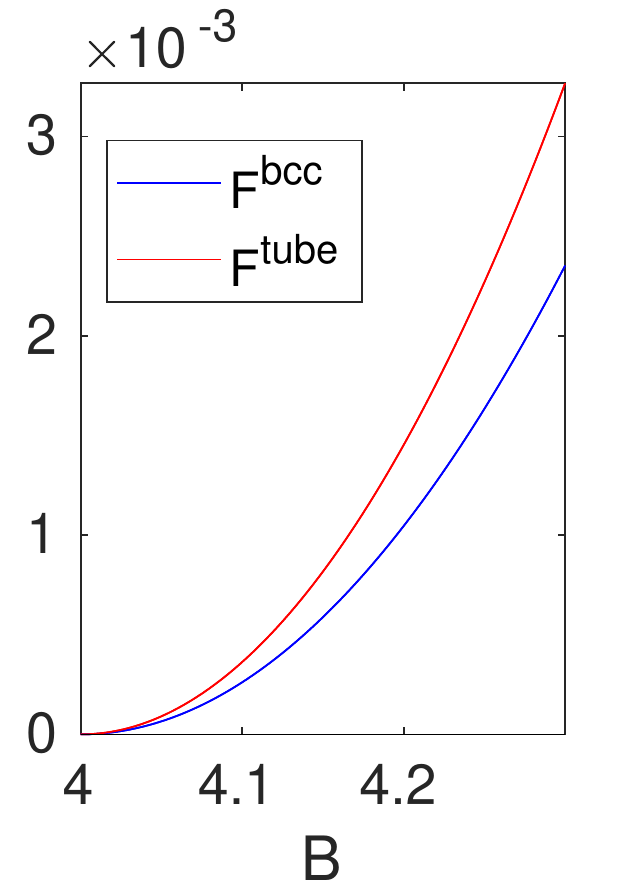}&
\hs{-0mm}\ig[width=0.25\tew]{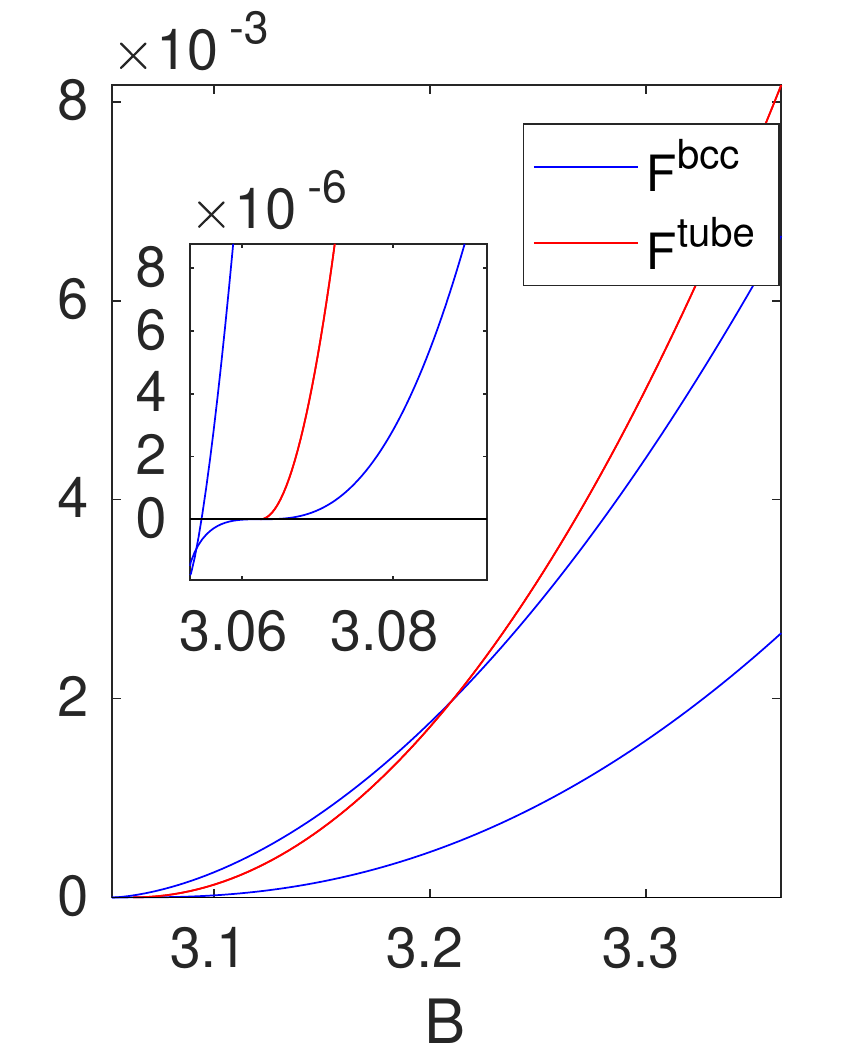}&\hs{-0mm}\ig[width=0.23\tew]{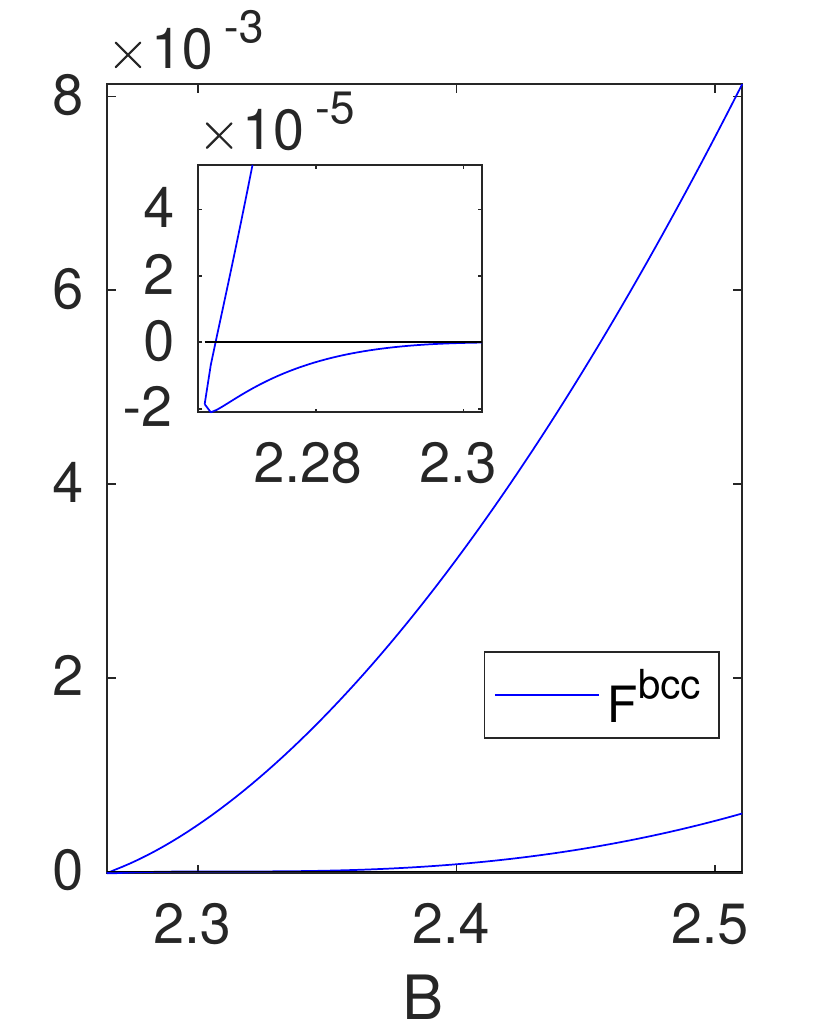}&
\hs{-0mm}\ig[width=0.25\tew]{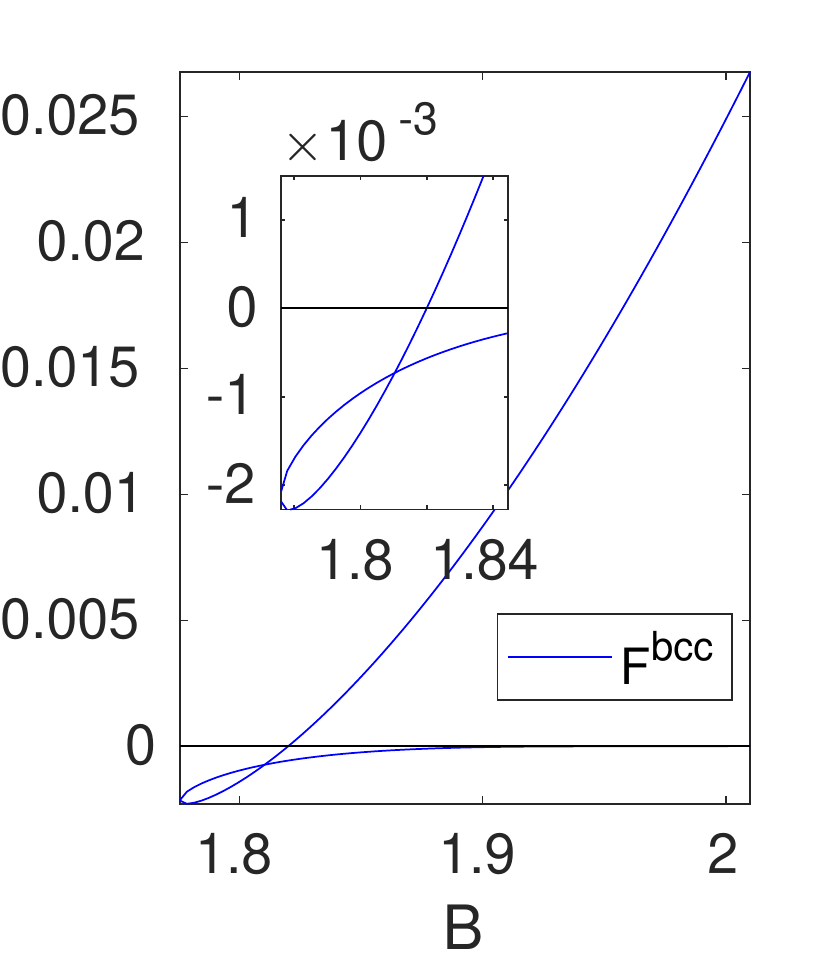}
\end{tabular}
}
\ece 

\vs{-5mm}
\caption{{\small Energy $F$ for the BCCs $\Abv_\pm$ and the tubes 
$\Atv$ (a,b) plotted as a function of $B$, where $\lam=\del(B-B_c)$, values 
of $R$ as indicated. The insets show a zoom near the primary bifurcation. 
The intersection of $F$ with $F=0$ defines the (approximate) Maxwell points 
$B_M$ from Table \ref{lctab}, and the intersections of $F$ in (b) the 
Maxwell point $\tilde{B}_M$. In (c,d) we omit $F(\Atv)$ because any 
reasonable approximation of the tubes here would need 5th order amplitude 
expansions. 
\label{hf4b}}}
\end{figure}

The amplitude system \reff{reda} also already contains the information 
to derive a necessary condition for fronts between BCCs and zero, or 
BCCs and tubes to exist on the level of the amplitude equations \reff{reda}.  
If we assume a slow $z$ dependence of the amplitudes $A_1,A_2$, then we 
can formally derive an extension of \reff{reda} to 
\huga{\label{GLred}
\barr{rl}
\dot A_1&=-d_4\pa_z^4 A_1+\lam A_1+aA_2^2+b(\frac 3 2 A_1^2+6A_2)A_1,\\
\dot A_2&=d_2\pa_z^2 A_2+\lam A_2+aA_1A_2+b(3A_1^2+\frac 9 2 A_2^2)A_2. 
\earr
}
The second order coefficient $d_2$ is determined as $d_2=-\frac 1 2 (n^{(2)}\cdot (0,0,\pa_k))^2\mu_1(k)$ where $n^{(j)}=\vk^{(j)}/\|\vk^{(j)}\|_2$, cf., 
e.g., \cite[\S4.6]{pismen06}. For the mode $A_1\er^{\ri \vk^{(1)}\cdot \vx}$ we have 
$n^{(1)}\cdot (0,0,\pa_k)=0$ and hence must expand the dispersion relation 
$\mu_1(k)$ to 4th order around $\vk^{(1)}$, yielding  $d_4=\frac 1 {4!} 
\pa_k^4\mu_1(k_c)$. 
\def\Ekin{E_{\text{kin}}}
The system \reff{GLred} has the conserved quantity $E=\Ekin+F$, i.e., 
$\frac{d}{dz}E(A(z))=0$, where 
\huga{\label{ekin}
\Ekin=d_4\left[\pa_z^3A_1\pa_zA_1-\frac 1 2 (\pa_z^2 A_1)^2\right]+d_2(\pa_zA_2)^2
}
can be considered as a kinetic energy, and 
\huga{\label{fdef}
F=\frac 1 2 \lam (A_1^2+2A_2^2)+2qA_1A_2^2+\frac\al 4 A_1^4+\beta A_1^2A_2^2
+\frac \ga 2 A_2^4
}
as a potential energy. Thus, a  
necessary condition for the existence of steady front solutions of \reff{GLred}, 
connecting, e.g., $\Abv_+$ at $z=-\infty$ with $\vec{A}=(0,0)$ 
at $z=\infty$, or $\Abv_+$ at $z=-\infty$ with $\Atv$ 
at $z=\infty$, is that the limit states (where $\pa_z=0$) 
have the same potential energy, i.e., 
\hual{
F(\Abv)=0&\text{ for a heteroclinic between $\Abv$ and $(0,0)$,}\\
F(\Abv)=F(\Atv)&\text{ for a heteroclinic between $\Abv$ and $\Atv$.}
}
These equalities only hold at specific points, the Maxwell points. As already 
said, for \reff{bruss1} we use $B$ as a bifurcation parameter, 
for fixed $R=0.75, R=0.52$ and $R=0.4$ (and, for completeness, $R=1$), yielding 
the coefficients $\del,q,\al,\beta,\ga$ from Table \ref{lctab}. 
In Fig.~\ref{hf4b} we plot, for these values, 
$F$ for the $\Abv$ and $\Atv$ branches as 
a function of $B$, which defines $\lam$ via $\lam=\del(B-B_c)$. 
This illustrates four main things, and in the following section we use 
these (formal) results from the amplitude system 
\reff{reda} as a guide to search for the associated solutions of the original system 
\reff{bruss1}: 
\bci 
\item For $R=1$, the bifurcations of the BCCs and tubes are supercritical, 
and the $F$--plots show that no fronts between any of $\Abv_+,\Atv$ and $\vec{A}=0$ can exist. Thus we also do not expect such fronts in the full system, 
at least not near onset where we expect the amplitude equations to make 
good predictions. 
\item For $R=0.75$, there is a 
Maxwell point $B_M$ near $B=3.054$ for a front between $\Abv_+$ and $(0,0)$. 
However, the subcriticality is very weak 
and thus we should expect finding the associated  fronts (if they exist) 
in the original system \reff{bruss1} to be a delicate and expensive task, 
cf.~Remark \ref{hexrem}(b). 
For $R=0.75$ there additionally exists a Maxwell point 
$\tilde{B}_M$ near $B=3.21$ for a front between $\Abv_+$ and $\Atv$, 
and thus the possibility of steady fronts between $\wbcc$ and $\wtube$ 
in this parameter regime. This prediction, in particular 
the quantitative value for $\tilde{B}_M$, should only be considered as 
a hint as we are relatively far from onset. 
\item For decreasing $R$, the fold $B_f(R)$ and the Maxwell 
points $B_M(R)$ move farther away from $B_c(R)$ (cf.~Table \ref{lctab}). 
Thus, if steady fronts between $\wbcc$ and $\wtube$ exist, then 
they should be easier to find at smaller $R$. 
\item On the other hand, at $R=0.52$ 
the pitchfork for $\Atv$ is subcritical, and we should not 
expect the associated branch to make any reasonable predictions 
away from onset. In particular, while the sub/vs supercritical 
branching behavior of $\wtube$ is correctly predicted, 
the $\wtube$ branch in \reff{bruss1} has a fold rather close to onset, and this 
behavior can only be resolved by 5th order amplitude equations, which we 
do not consider here. 
\eci 

\section{Results for the full system}
To illustrate/corroborate some results from the amplitude formalism, and to find 
snaking branches for \reff{bruss1}, we use \pdep\ \cite{p2pure, p2phome}. 
The approach is 
motivated by (and the results essentially similar to) the results on 
snaking branches for 1D and 2D problems in \cite{uwsnak14, pftuta}, but as already noted in the Introduction, the 3D case does present a number of significant 
numerical challenges. Besides the obvious issue of higher numerical costs 
due to more degrees of freedom (DoF) 
in the (discretized) 3D case, these challenges 
mainly include  
the branch switching at branch points of high multiplicity, 
and, in particular over non-small small domains, 
problems with undesired 'branch jumping' due to many solution 
branches close to each other (more than in 1D and 2D). 
See \cite[\S3]{pftuta} for details on how we deal with these problems. 
Here we only remark that: 
\bci 
\item The branch switching proceeds by (numerically) deriving and solving 
the pertinent algebraic bifurcation equations, which are essentially 
equivalent to the amplitude equations. No specific knowledge of 
the structure of the bifurcation problem is needed for this, 
but the user can (and should) use the symmetries to make a selection of branches 
to be continued. 
\item To have reliable and fast numerics we stick to rather small 
domains; in particular, for fronts we extend small $x$--$y$--squares 
in $z$--direction, i.e., choose long and slender bars. 
The typical number of DoF in the 
results below is on the order of $10^5$, and the residual tolerance is 
\huga{\label{resi}
\res:=\|G(u)\|_\infty\stackrel!<10^{-8}, 
} 
where $G(u)$ is the FEM discretization of the right hand side of \reff{bruss1}. Typical runtimes for continuation of branches on an I7 
laptop are, e.g., about 10-20 min for 50 points, including 
the stability computation. \huc{In detail, e.g., for the cubic domains in 
Fig.~\ref{s052} we have 83.000 mesh points, which on a rectangular 
grid would correspond to 
about 44 points in each spatial direction. However, for symmetry reasons 
a staggered grid of 'symmetry type 1' (see \cite[\S3.4]{pftuta}) is 
used. A zoom of a typical mesh for the long and slender bars used to 
compute fronts is given in Fig.\ref{s052}(h). }
\item \huc{To check accuray we essentially 
recomputed selected solutions on 
finer grids, typically with double the DoF. When first interpolating 
\def\ui{u^{\text{interp}}}\def\ufi{u^{\text{fine}}}
a given solution to a solution $\ui$ on the finer grid and then 
using a Newton--loop to obtain a solution $\ufi$ fulfilling \reff{resi} 
on that grid, then in all cases 
$\|\ui-\ufi\|_\infty<0.005$. Moreover, a few continuation steps 
on the finer grid yield the same behaviour as on the original grid. 
Additionally, we used adaptive mesh--refinement on selected 
solutions (see Fig.~\ref{s052} for an example), with 
$\|u-u^{{\rm refined}}\|_\infty<0.005$, and without 
visible changes in the solution structure.  
Thus we believe that our meshes are sufficiently fine that the 
numerical solutions are converged and reflect the true PDE behaviour. }
\eci

\subsection{Fronts between BCCs and $u^*$}\label{b20sec}
\begin{figure}[ht]
\bce
{\small
\begin{tabular}{llll}
(a)&(b) {\tt BCC/pt22}&(c) {\tt BCC/pt22}&(d) {\tt tubes/pt10}\\
\ig[width=0.28\tew,height=45mm]{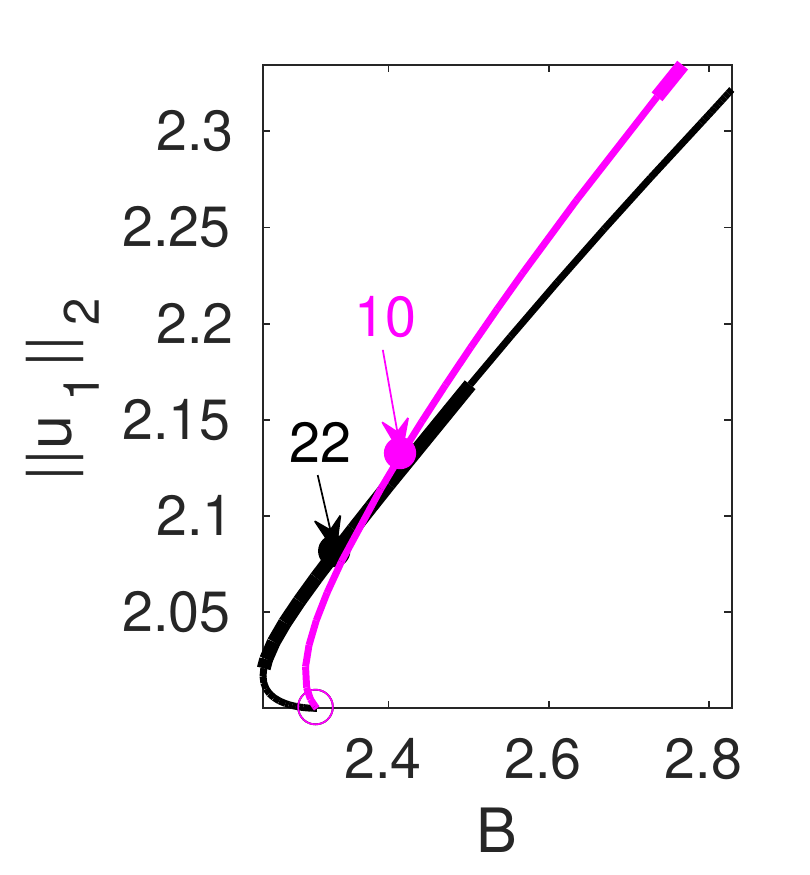}&
\raisebox{3mm}{\ig[width=0.24\tew]{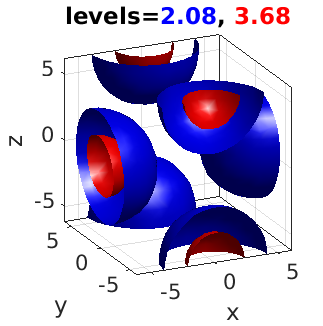}}&
\ig[width=0.2\tew]{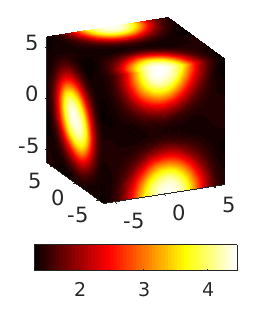}&
\ig[width=0.2\tew]{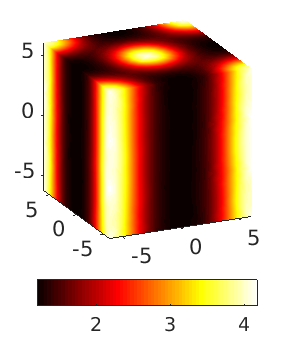}
\end{tabular}\\
\begin{tabular}{lllll}
(e)&(f)&(g)&(h)&(i)\\
\hs{-5mm}\raisebox{30mm}{
\begin{tabular}{l}\ig[width=0.21\tew]{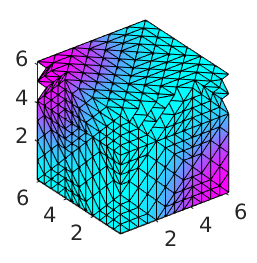}\\[-2mm]
\ig[width=0.21\tew]{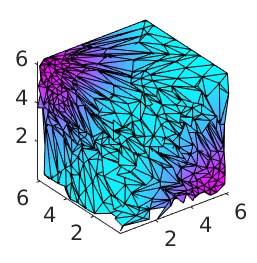}\end{tabular}}
&\ig[width=0.4\tew,height=60mm]{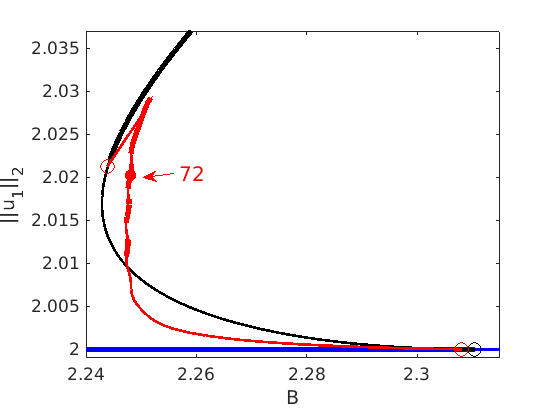}
&\hs{-3mm}\raisebox{-4mm}{\ig[width=0.11\tew]{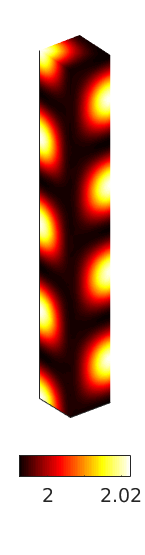}}
&\raisebox{-4mm}{\ig[width=0.11\tew]{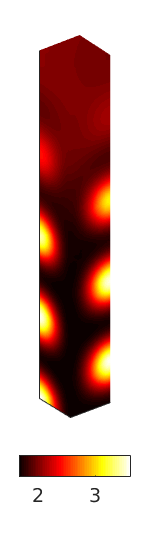}}
&\hs{-3mm}\raisebox{8mm}{\ig[width=0.16\tew]{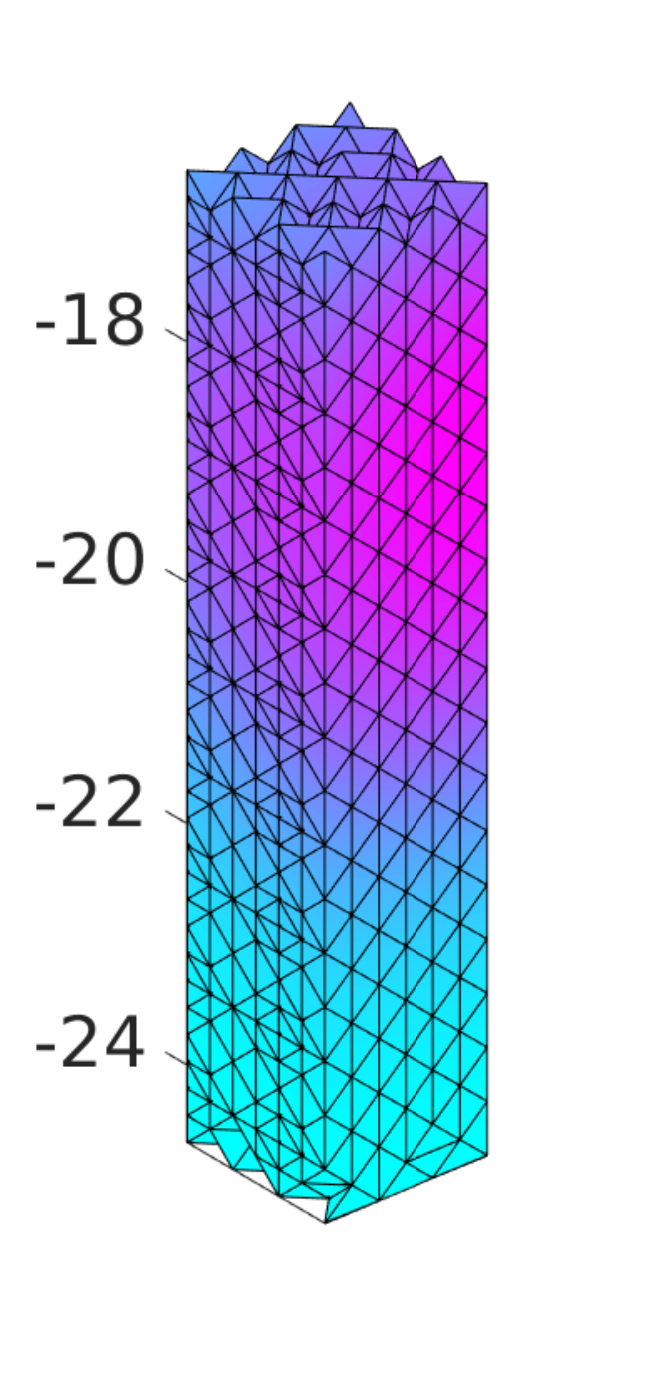}}
\end{tabular}
}
\ece

\vs{-5mm}
\caption{{\small $R=0.52$. (a) Branches of the BCCs (black) and tubes (magenta) 
for \reff{bruss1} over 
the cube $\Omega=(-l,l)^3$ with $l{=}\sqrt{2}\pi/k_c$ and Neumann boundary conditions. The norm $\|u_1\|_2$ is the normalized $L^2$ norm 
$\left(\frac 1{|\Om|}\int_\Om u_1(x)\dd x\right)^{1/2}$. 
(b)--(d) sample plots from (a). (b) shows isosurfaces of $u_1$, while 
(c,d) show $u_1$ on the surface on the domain. 
(e) Illustration of meshes (uniform and adapted), 
plotted over 1/8th of the computational domain. 
(f) Branches of the homogeneous solution (blue), the BCCs (black), and localized BCCs (red) over the domain $\Omega=(-l_x,l_x)\times (-l_y,l_y)\times (-l_z,l_z)$ with $l_x{=}l_y{=}\sqrt{2}\pi/(2 k_c)$ and $l_z{=}4\sqrt{2}\pi/k_c$. 
(g) $u_1$ for BCCs near onset. (h) $u_1$ for point 72 in (e), showing a front 
between $u=u^*$ and the {\tt BCCs}. (i) illustration of the uniform 
mesh used in (f--h) by replotting solution (h) on a $1/96$--part of $\Om$ 
at the bottom front right of $\Om$.  \label{s052}}}
\end{figure}

First we seek fronts between BCCs and $u^*$. Following Remark \ref{hexrem}(b), 
this should be easier for smaller $R$ than for $R$ close to 1. 
In Fig.~\ref{s052} we start with $R=0.52$ which is roughly the largest value for 
which we find a snaking branch of a front between BCCs and $u^*$ on a domain 
as in Fig.\ref{s052}(f--h). 
We come back to $R=0.75$ in \S\ref{b2tsec} 
where we consider the bistability range of BCCs and tubes. 

Figure \ref{s052}(a) shows the bifurcation diagram of BCCs and tubes on a small cube 
(8 times the minimal domain, i.e., $\Om=(-l,l)^3$ with $l=\sqrt{2}\pi/k_c$), 
and (b)--(d) shows sample plots. The BCC branch qualitatively (and also 
quantitatively) agrees with the 
predictions from \reff{reda}. The tubes (magenta) bifurcate subcritically (as predicted), but 
in the full system have a fold close to the bifurcation and hence can 
only be approximated by the amplitude system close to onset.  
\huc{In (e) we illustrate adaptive mesh refinement from a solution 
as in (b), for graphical reasons showing 1/8 of the computational 
domain and starting with 
a relatively coarse uniform mesh with $n_p=10351$ grid points  
(half the number of mesh points in each direction compared to the 
computations in (a)--(d)), 
and then adapting to $n_p=17258$. See \cite{trulletuta} 
for the setup of 3D mesh adaptation in \pdep, based on the package 
{\tt trullekrul} \cite{KEJ17}, and \cite[\S3.4]{pftuta} for examples  
of the usage for localized 3D patterns in the Swift--Hohenberg equation.}

\huc{In (f)--(h) we focus on the subcritical range on a long and slender bar 
$\Omega=(-l_x,l_x)\times (-l_y,l_y)\times (-l_z,l_z)$ with $l_x=l_y=\sqrt{2}\pi/(2 k_c)$ and $l_z=4\sqrt{2}\pi/k_c$, and with $1.8\times 10^5$ DoF. 
The snaking red branch bifurcates from the BCCs (black branch) close 
to onset, and corresponds to a front between BCCs and $u^*$. In the snaking 
region around $B=2.25$, which is reasonably 
close to the Maxwell point prediction $B_M=2.266$, it alternates between stable and unstable parts, 
and in each pair of folds an additional layer of BCCs is added. 
In (i) we illustrate the (uniform) meshing in (f)--(h) on the subdomain 
$(l_x/2,l_x)\times(l_y/2,l_y)\times(-l_z,-2l_z/3)\subset\Om$ corresponding 
to $\frac 1 4 \frac 1 4 \frac 1 6=\frac 1 {96}$ of the full domain 
at the bottom front right.}

\begin{figure}[h]
\bce{\small
\begin{tabular}{llll}
(a)&(b) pt1&(c) pt37&(d) pt90 \\
\ig[width=0.5\tew]{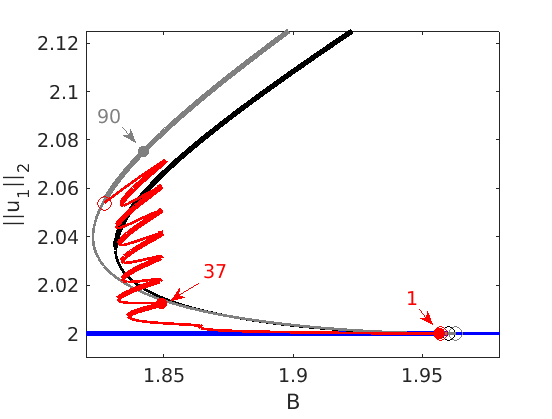}&
\ig[width=0.1\tew]{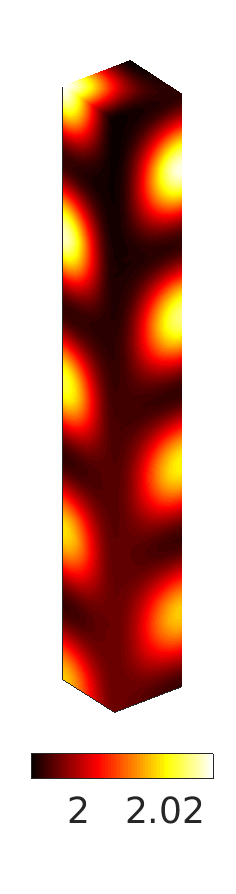}&
\ig[width=0.1\tew]{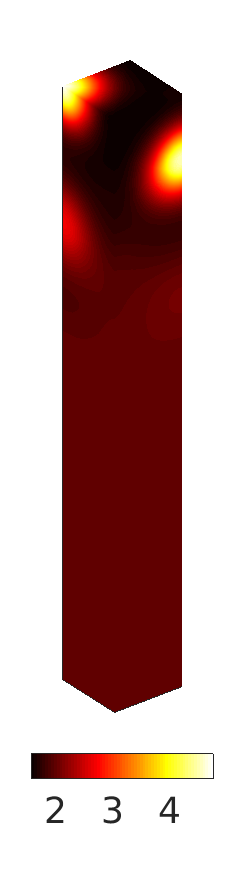}&
\ig[width=0.1\tew]{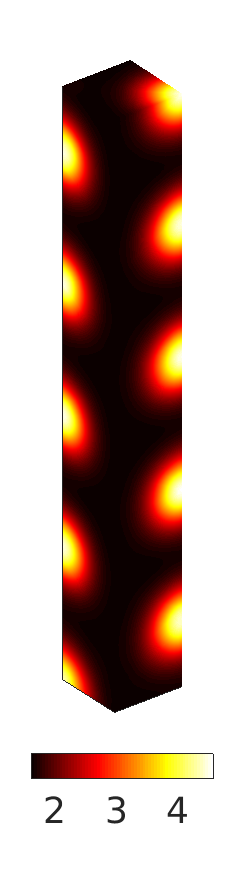}
\end{tabular}}
\ece 

\vs{-4mm}
\caption{{\small $R=0.4$. (a) Homogeneous branch (blue), BCCs (black), slightly distorted balls (grey), and localized BCCs (red), $\Om=(-l,l)^2\times(-4l,4l)$ with 
Neumann BC.   The snaking red branch bifurcates from the BCCs shortly 
after the primary bifurcation and reconnects to the distorted cubes. 
(b), (c), (d) shows sample solutions from the labels in (a).}}\label{s04}
\end{figure}

To increase the narrow snaking region in Fig.~\ref{s052}, 
we lower $R$ further to $R=0.4$ in  Fig.~\ref{s04}. The branch of localized BCCs bifurcates from the BCC branch near onset as before, but is now to the 
right of the Maxwell point prediction $B_M=1.82$. 
However, during the snaking the localized BCCs significantly 
change their wave lengths 
in $z$ direction and terminate in a pitchfork bifurcation on a branch 
corresponding to cubes of the form (modulo a phase shift in $x,y$) 
\def\kati{\tilde{\kap}}
\huga{\label{dBCC}
w= 4A^*[ \cos(\kappa (x+y))+\cos(\kappa(x-y))+\cos(\kappa y+\kati z)]\Phi+\text{h.o.t.}
}
with $\kappa=k_c/\sqrt{2}$ and $\kati=9\kap/8$. 
Such shifts to patterns with slightly different $|\vk|$ (sideband patterns), 
are also known from 1D and 2D, cf., e.g., \cite{uwsnak14}. 
\huc{They are analyzed (aided by numerics) in detail for the 1D quadratic--cubic 
Swift--Hohenberg equation in \cite{BBKM08} on a finite domain, 
including a relation to the Eckhaus instabilities  
of the periodic branches, 
and the phenomena explained there also occur here: The termination of the 
snake at the upper end where the pattern almost fills the domain 
depends critically on the domain size and 
the strength of the subcriticality, i.e., the snaking width. 
For stronger subcriticality, the local 
wave number $k_{\text{loc}}$ of the patterns in the snake 
varies more strongly, and then typically the snake bifurcating 
from the primary pattern ($|\vk|=k_c$) tends to terminate on the 
branch of periodic patterns which is most subcritical, i.e., here 
the grey branch, which has the smallest $B$ value in its fold. }

\huc{This also indicates that for small $R$ 
we should not expect the primary BCCs to be the ``most stable'' 
pattern. Instead, the stability range of 
``distorted'' BCCs like in Fig.~\ref{s04}(d) 
may extend to significantly lower $B$ than that of the primary BCCs.  
In Fig.~\ref{s04} 
(like in Fig.~\ref{s052} and in similar figures below) we have 
a quasi 1D situation due to the small domain size in $x$ and $y$. 
Moreover, also the domain size in $z$ is not really large, and together with 
the Neumann BCs this restricts the allowed wave vectors to a still rather 
small set. Over larger domains, the variety of (stable) patterns 
and associated possible snaking away from onset 
rather quickly becomes excessive, and for instance the stability 
ranges of different distorted BCCs will deserve a dedicated study. 
}

\subsection{Fronts between BCCs and tubes}\label{b2tsec}
\begin{figure}[ht]
\bce{\small 
\begin{tabular}{lllllll}
(a1)&(b)&(c1)&(d) pt1&(e) pt40&(f) pt65&(g) pt30\\
\hs{-5mm}\raisebox{40mm}{\begin{tabular}{l}
\ig[width=0.18\tew,height=40mm]{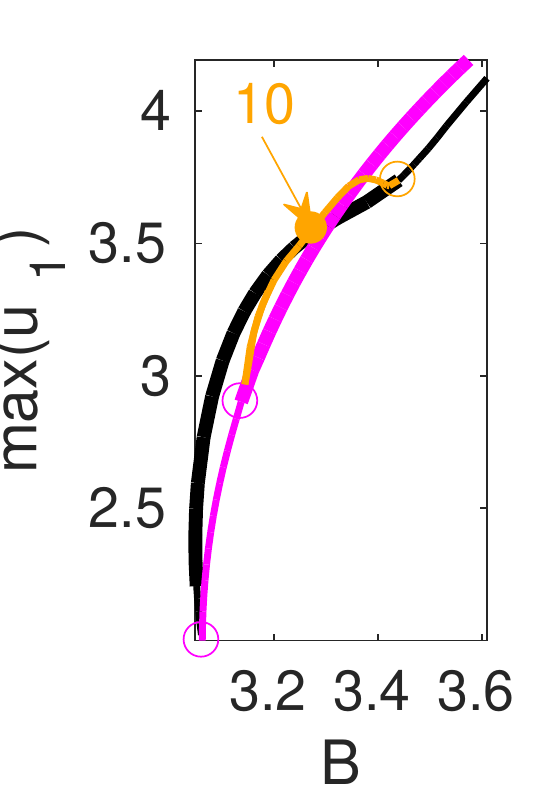}\\
(a2)\\\ig[width=0.14\tew]{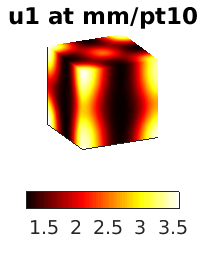}
\end{tabular}}
&\ig[width=0.09\tew]{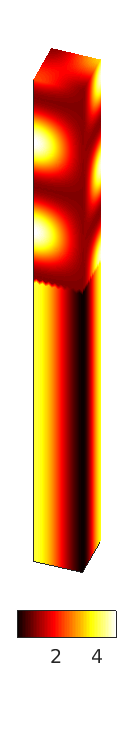}
&\raisebox{40mm}{\begin{tabular}{l}
\ig[width=0.23\tew,height=25mm]{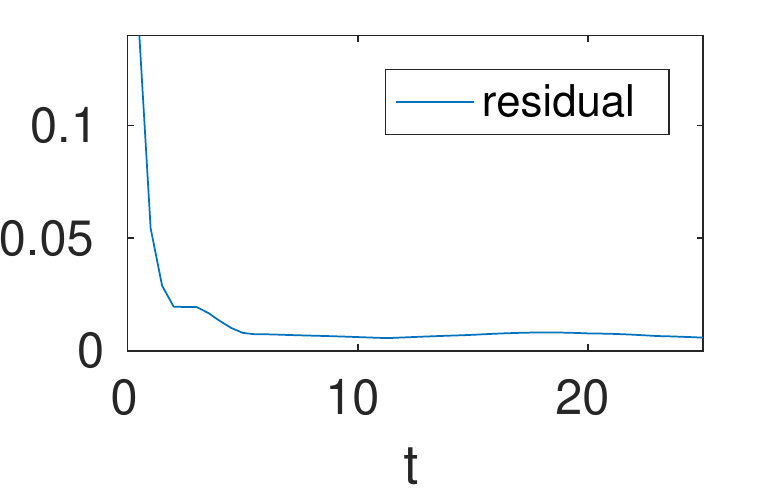}\\
(c2)\\
\hs{-5mm}\ig[width=0.29\tew,height=44mm]{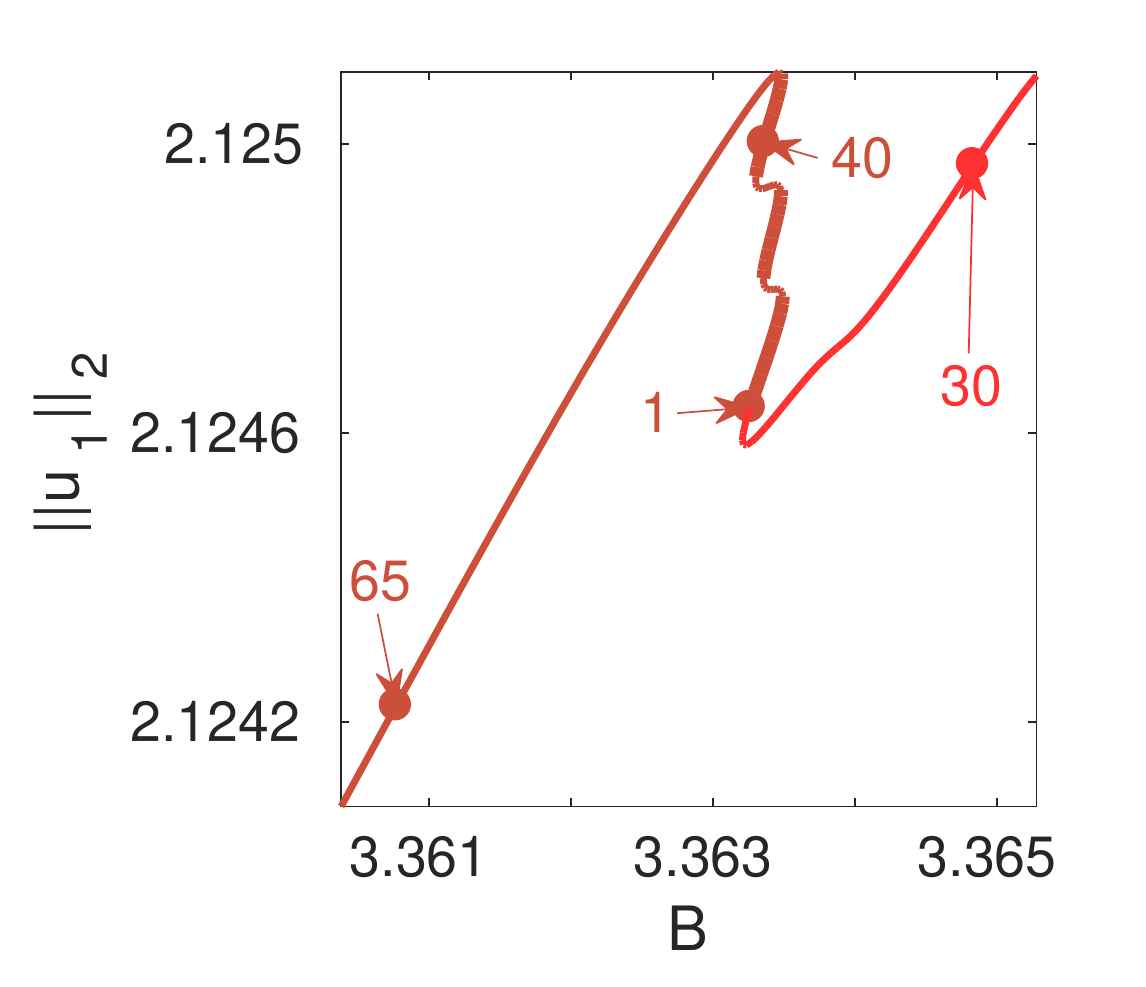}
\end{tabular}}
&
\hs{-3mm}\ig[width=0.09\tew]{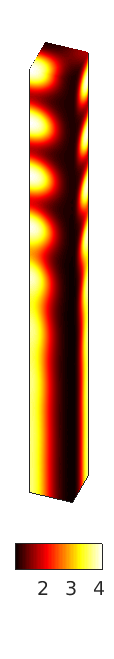}
&\hs{-3mm}\ig[width=0.09\tew]{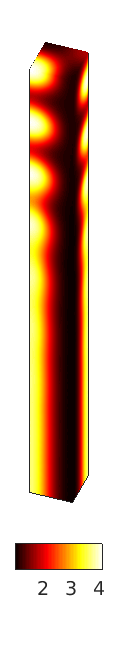}
&\hs{-3mm}\ig[width=0.115\tew]{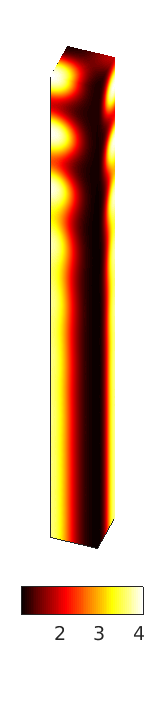}
&\hs{-3mm}\ig[width=0.09\tew]{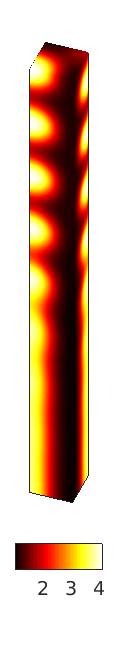}
\end{tabular}
}
\ece 

\vs{-8mm}
\caption{{\small $R=0.75$. (a1) BD of (primary) BCCs (black), tubes (magenta), and mixed modes (orange) over the cube $\Om=(-l,l)^3$, $l=\sqrt{2}\pi/k_c$, cf.~Fig.~\ref{hf4}. Here 
we plot $\max(u_1)$ because in the 'standard norm' $\|u_1\|$ the three 
branches are close to each other in the bistable range $3.17<B<3.42$. 
(a2) Sample plot on the mixed mode branch.  
(b) Initial condition \reff{ics} for the DNS to obtain a tube--to--BCC front;  
$\Om=(-l,l)^2\times (-l_z,l_z)$, where $l=\pi/(\sqrt{2}k_c)$ and 
$l_z=12l$. 
(c1) Initial evolution of $\res(u(t))$ for the DNS. (c2) 
BD of a narrow snake of fronts between (distorted) BCCs and tubes. 
The starting point $u_f$ (pt1, see (d))) on that snake was obtained from DNS 
followed by a Newton loop. 
(e)--(g): Sample plots from the continuation into the direction of 
tubes (e,f) and spots (g).\label{hf1}}}
\end{figure}

In \S\ref{maxsec} we explained that for $R=0.75$ the amplitude equations 
\reff{reda} also predict mixed mode branches (orange branch in 
Fig.~\ref{hf4}(a)), which suggests the existence of fronts between 
BCCs and tubes, cf.~Fig.\ref{hf4b}(b).
However, these occur at $\CO(1)$ distance from onset, and hence such 
predictions should be taken with caution. 
In Fig.~\ref{hf1}(a) we show the BCCs (black), tubes
(magenta) and mixed modes (orange) for \reff{bruss1} with $R=0.75$
over the cube $\Om=(-l,l)^3$, $l=\sqrt{2}\pi/k_c$. This confirms the
predictions from Fig.~\ref{hf4} over this small domain, and we may
extend these periodic patterns over the boundaries to obtain the same
patterns and branches over larger domains. However, it turns out
that even the reliable continuation of the BCC branch to $B-B_c=\CO(1)$ 
over extended
domains is a delicate task, and requires fine meshes and strict
settings for the algorithm {\tt pmcont} designed to mitigate undesired
branch switching, see \cite[\S3.3]{pftuta}. See also the remarks at the end 
of \S\ref{b20sec}.

Therefore we proceed differently to explore the range $B\approx \tilde{B}_M$ 
for fronts 
between BCCs and tubes over long and slender bars, aiming to find 
snaking branches of fronts between BCCs and tubes. Figure \ref{hf1}(b) shows 
an initial condition (IC) of the form 
\huga{\label{ics} u_1(x)=\left\{\barr{ll}u_{{\rm BCC},1}& z>5\\
u_{{\rm tube},1}& z\le 5\earr\right., 
}
 composed of the 
primary BCCs above the interface at $z=5$, and tubes below, while 
$u_2$ is simply set to the homogeneous value $u_2=u_2^*=B/A$. 
The domain is $\Om=(-l,l)^2\times (-l_z,l_z)$, where $l=\pi/(\sqrt{2}k_c)$ and 
$l_z=12l$. Additionally, we choose rather carefully (see below) the value 
$B=3.3633$. Starting with this initial condition, direct numerical simulation 
(DNS) 
slowly decreases the residual $\res(u(t))$ defined in \reff{resi}, see (c1). 
However, this decrease is in general not monotonous, and the BCCs actually 
change their wave vector.  Nevertheless, after the transient (at, e.g., $t=25$) 
we can 
run a Newton loop on the stationary problem, 
and converge to the solution $u_f$ illustrated in (d). 
This is a (stable, as it is essentially obtained from DNS) stationary 
front between BCCs on top and tubes at the bottom, but similarly to 
Fig.~\ref{s04}(d), the BCCs are clearly 
not the primary BCCs belonging to \reff{wav}, but (rather strongly) 
distorted, i.e., of the 
form \reff{dBCC} with $\kati=1.5\kappa$. 
Next we continue $u_f$ in $B$, and obtain the (narrow and short) snake 
shown in (c2). In one direction (brown part), the spots recede 
(sample plots (e,f)) as the parameter varies, and 
in the other direction (red part), the spots expand (sample plot (g)). 
In both directions, the branch eventually reconnects to the mixed mode 
branch between the tubes and the BCCs with $\kati=1.5\kappa$. 

\begin{figure}[ht]
\bce{\small 
\begin{tabular}{llll}
(a)&(b)&(c)&(d)\\
\hs{-6mm}\raisebox{45mm}{\begin{tabular}{l}
\ig[width=0.22\tew]{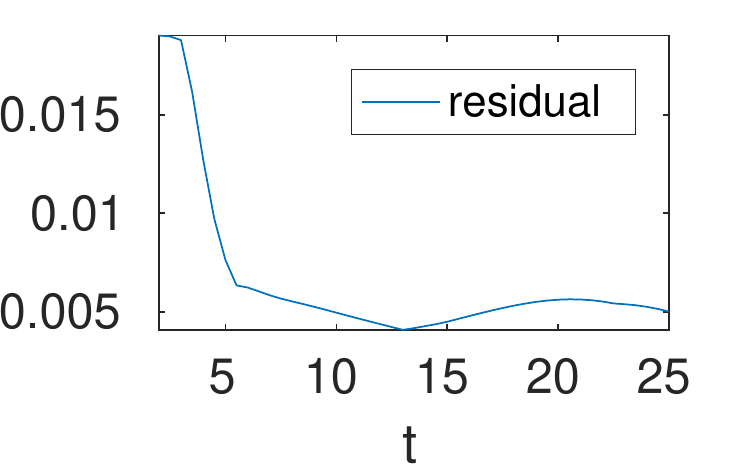}\\
\ig[width=0.22\tew]{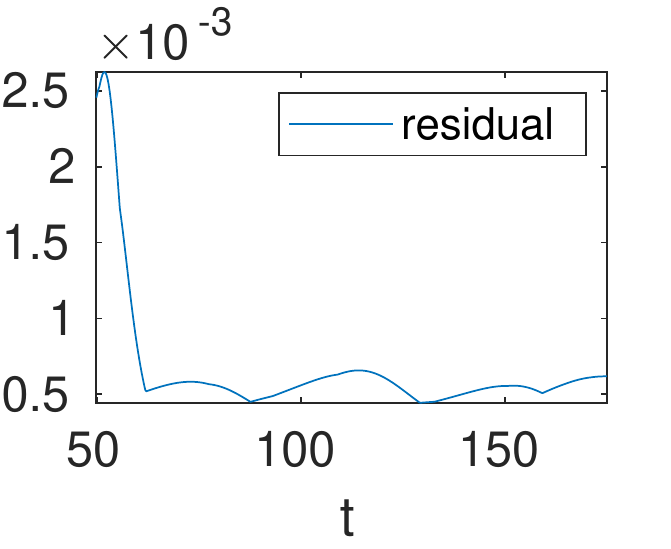}
\end{tabular}}
&
\ig[width=0.09\tew]{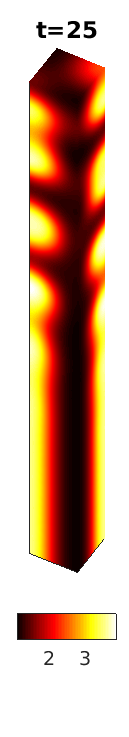}
\ig[width=0.09\tew]{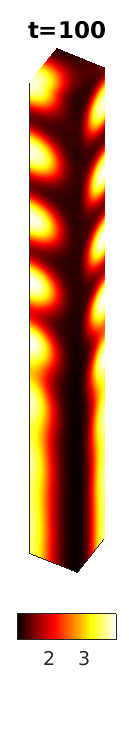}
\ig[width=0.09\tew]{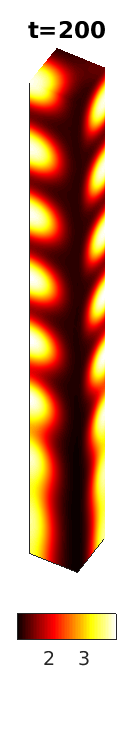}
&\hs{-3mm}\raisebox{45mm}{\begin{tabular}{l}
\ig[width=0.22\tew]{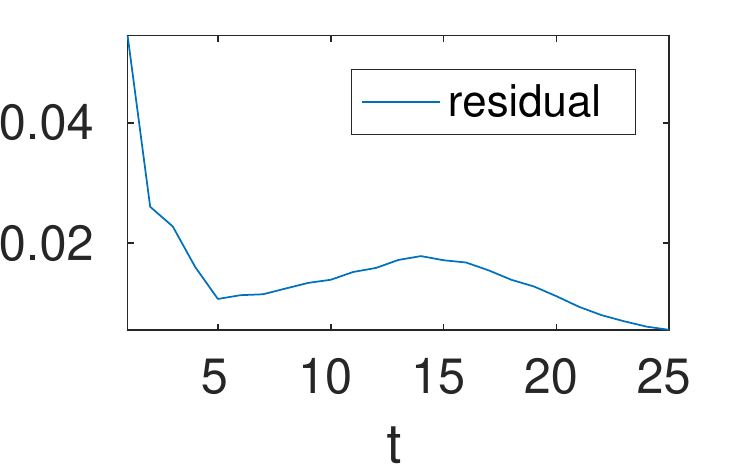}\\
\ig[width=0.22\tew]{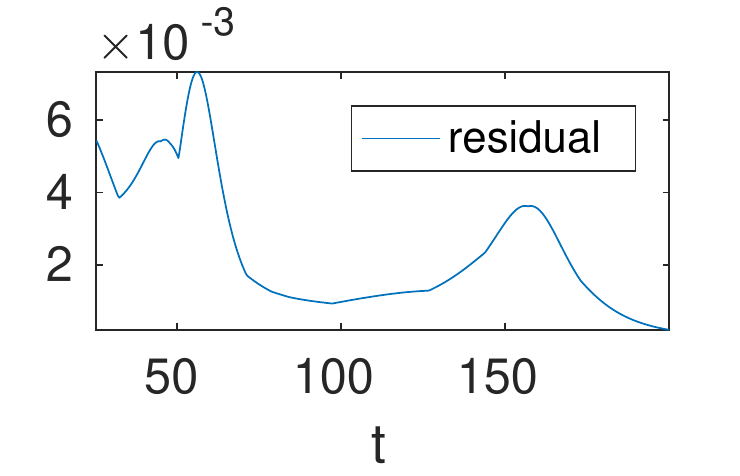}
\end{tabular}}
&
\ig[width=0.09\tew]{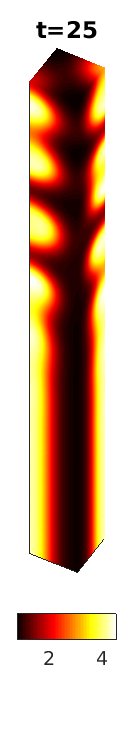}
\ig[width=0.09\tew]{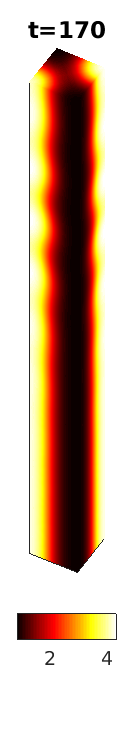}
\end{tabular}
}
\ece 

\vs{-10mm}
\caption{{\small $R=0.75$. Illustration of DNS for choices of $B$ outside 
the snaking range. For $B=3.6$ (right of the snake from Fig.~\ref{hf1}) the solution converges to a BCC solution 
in a long transient with stick-slip motion  (a,b).  For $B=3.3$ (left of the snake) the tubes win (c,d).  \label{hf2}}}
\end{figure}

The snake in Fig.~\ref{hf1} is rather narrow, and the starting point was 
obtained by a careful choice of $B$ for the DNS. In Fig.~\ref{hf2} 
we illustrate the ``typical'' behavior of DNS for ICs of the form \reff{ics}, 
which also explains the idea how to find $B$ for Fig.~\ref{hf1}. 
We use the same domain and IC as in Fig.~\ref{hf1}. 
For $B=3.6$, in (a,b), the initial dynamics 
is very similar to that in Fig.~\ref{hf1}, 
i.e., the solution evolves towards a (distorted) BCC-tubes fronts. 
However, once the solution is ``near the snake'' from Fig.~\ref{hf1}, 
the BCC part continues to grow in time. 
% roughly similar to 
%stick--slip motion, as can be seen in the time series in (a) (and (c)). 
\huc{If we were close enough to the snake of steady fronts, on a 
sufficiently large domain, then we would expect ``stick--slip'' motion. 
See, e.g., \cite[\S III.B]{burke} where this is analyzed semi--analytically 
for fronts between patterns and the trivial 
solution outside the pinning (snaking) region in the (1D) Swift--Hohenberg 
equation. 
The motion is slow when the moving front passes near a steady solution 
at a fold of the corresponding snake, and afterwards moves quickly to 
near the next fold. The transition time from one fold to the next 
can be formally derived to be $\CO(\del^{-1/2})$ 
with $\del$ the distance from the snaking region, and the associated 
full formula shows 
excellent agreement with the numerics in \cite{burke}. See also 
\cite{DL19,DL20} for numerical analysis of the depinning of fronts 
in the planar Swift--Hohenberg equation. 
Here, the domain is not quite long enough, and for clarity we chose $B$ not 
very close to the snake, and thus we do not really see 
the stick--slip effect, but just roughly 
periodic variations in the residual, and for a rather short transient.}

If we use the solution from $t=200$ to start a Newton loop for the steady problem, then this gives convergence to the (distorted) BCC solution. Alternatively, continuing the DNS we also 
converge to this BCC after a very long transient. 
On the other hand, in (c,d) we choose $B=3.3$ to the left 
of the snake, and obtain convergence to the tubes, 
and the same happens (faster) at the Maxwell point 
prediction $\tilde{B}_M\approx 3.21$. 
If snaking branches containing stable steady fronts exist, then such simulations 
give a hint for the right parameters to find them, and that is how we found 
the $B$ value for Fig.~\ref{hf1} with some trial and error. 
In particular, the Maxwell point prediction $\tilde{B}_M\approx 3.21$ from 
Table \ref{lctab} was rather ``far off''. A certain deviation was expected 
a priori as we are at $\CO(1)$ distance from criticality. Additionally, and 
a posteriori, we see that $\tilde{B}_M\approx 3.21$ was irrelevant as it is 
the Maxwell point prediction for fronts between BCCs and tubes, and not 
the $\tilde{\kap}=1.5\kap$ distorted BCCs and tubes obtained in the DNS. 
We can, e.g., 
a posteriori change the $z$ wave number of the BCCs in the IC to start closer 
to $u_f$, which then also allows to directly go to $u_f$ by a 
Newton loop for the steady problem. However, Fig.~\ref{hf1} 
illustrates that very good initial guesses are often not necessary, and instead rather poor initial guesses can be first improved by DNS. 

In summary, the predictions from \S\ref{maxsec} are useful as 
they motivate the search for fronts and give hints for good parameter regimes. 
Of course, finding such fronts via DNS needs 
the existence of steady localized patterns of the desired form, and is easier and more robust if the (desired) snake is wide. 
On the other hand, in pattern forming systems such as \reff{bruss1} we may 
expect a (large) variety of (stable) steady patterns far from onset, and 
this increases the chances to converge towards {\em some} localized patterns. 
One more example is given in the next section, where at $R=1$ 
lamellas enter the game. 

\subsection{$R=1$: The comeback of the Lamellas}\label{lamsec}
\begin{figure}[ht]
\bce{\small 
\begin{tabular}{llll}
(a)&(b)&(c)&(d)\\
\hs{-5mm}\raisebox{10mm}{\ig[width=0.19\tew]{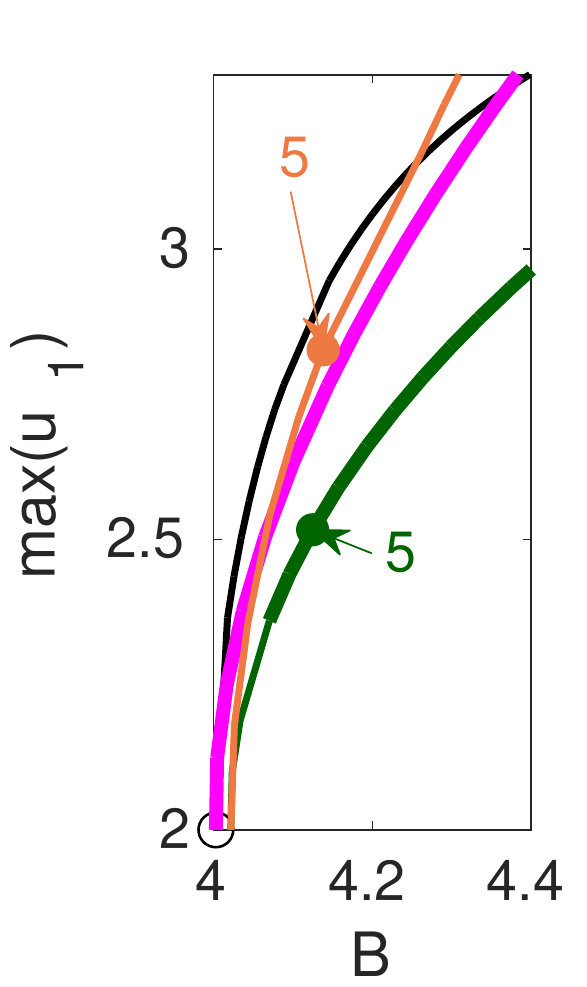}}
&\raisebox{10mm}{\ig[width=0.1\tew]{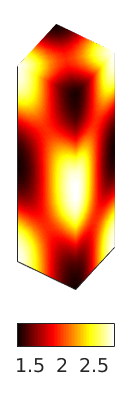}
%\raisebox{5mm}{
\ig[width=0.1\tew]{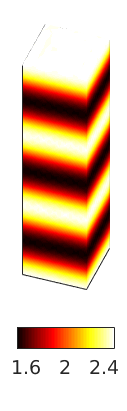}}
&
\raisebox{35mm}{\begin{tabular}{l}
\ig[width=0.24\tew]{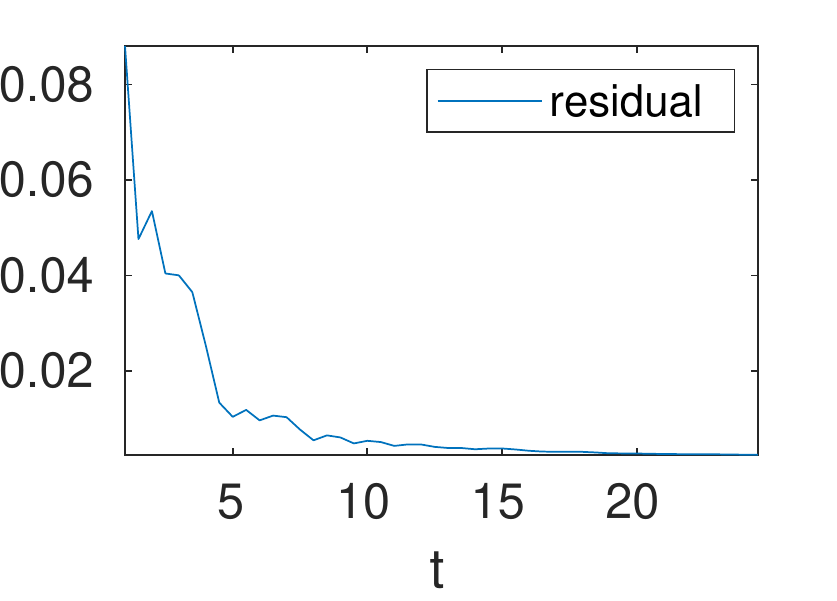}\\
\ig[width=0.24\tew]{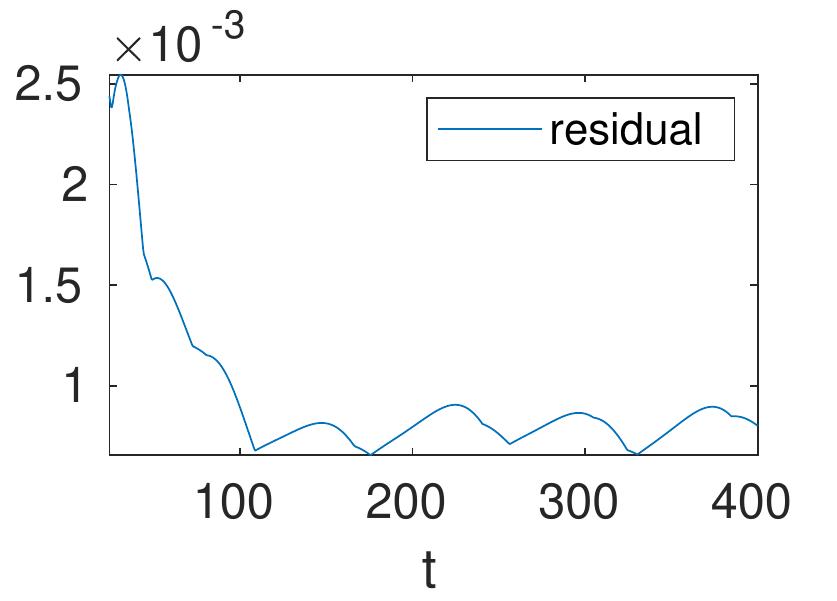}
\end{tabular}}
&
\hs{-4mm}\ig[width=0.1\tew]{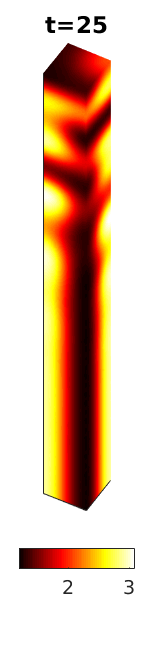}
\ig[width=0.1\tew]{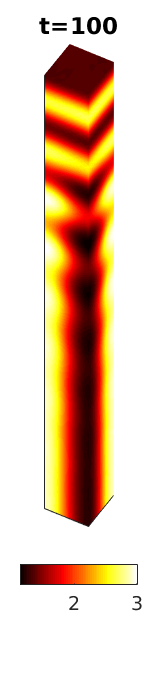}
%\ig[width=0.09\tew]{hupi/l1-200}
%\ig[width=0.09\tew]{hupi/l1-300}
\ig[width=0.1\tew]{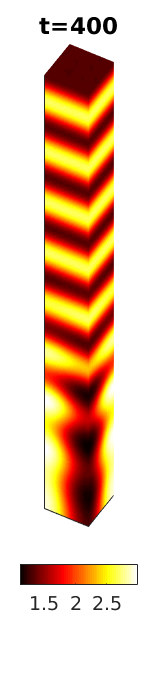}
\end{tabular}
}
\ece 

\vs{-10mm}
\caption{{\small $R=1$, hence $k_c=1$.  
(a) Bifurcation diagram of BCCs (black) and tubes (magenta), over the domain 
$\Om=(-l_x,l_x)^2\times (-l_z,l_z)$, $l_x=\pi/\sqrt{2}$, $l_z=4l_x$, 
including the next two branches on this domain, with example plots of 
the 5th points in (b), the  
$1.5k_c$ lamellas in green. The tubes are stable 
throughout, and the lamellas 
are stable for $B>B_b\approx 4.06$, and this remains true over longer bars, 
i.e., $l_z=ml_x$ with $m\ge 5$.  
(c,d) DNS at $B=4.2$ with an initial condition as in \reff{ics}, $\Om=(-l_x,l_x)^2\times(-l_z,l_z)$, $l_z=12 l_x$. Evolution to a moving front between 
lamellas and tubes. 
  \label{hf3}}}
\end{figure}

In Fig.~\ref{hf3} we illustrate some typical results for \reff{bruss1} at $R=1$.  In (a) we show the BD over a short bar. 
The BCCs and tubes now both bifurcate in supercritical 
pitchforks, with the tubes stable. Additionally, we show the next 
two bifurcating branches. The orange branch consists of elongated BCCs, 
and the green branch are $\kap=1.5k_c$ lamellas 
\def\wlam{w_{{\rm lam}}}
\huga{\label{lf1} 
\wlam\sim\cos(\kappa z)\text{ with }\kap=1.5 k_c, 
}
which we did not consider in \S\ref{afsec} as they do not bifurcate at the 
primary bifurcation, but (on the given domain) at the third bifurcation 
point from the $u=u^*$ branch. These lamellas become stable at 
$B=B_l\approx 4.06$, on this domain, but similarly also on much longer domains.  

Thus, we now have a bistable range between 
tubes and lamellas, and the lamellas turn out to play a crucial role 
in the DNS, as illustrated in (c,d). 
We set $B=4.2$, and again use an  
initial condition of type \reff{ics}. Though there are no lamellas in the 
initial condition, the solution initially 
(till $t=100$, say), relaxes 
to a 'double--front' from lamellas to tubes with a distinct 'cubes-like' 
interface in between. 
This front then propagates downwards in a roughly periodic 
fashion (see the lower 
time series in (c)) but with essentially fixed shape.% 
\footnote{The fixed shape appears to be another effect of locking due to 
the periodic pattern, as for double fronts between homogeneous states 
one would generically expect the middle state to expand or shrink. 
See, e.g., \cite{CsMi99} for another striking 
example of such double-fronts, namely a 'roll belt' ahead of hexagons 
invading the zero solution in a damped Kuramoto-Sivashinsky equation.} 
 A similar behavior occurs at other values of $B$ ($B>B_l$) and other 
initial conditions, i.e., for $R=1$ the lamellas always win on 
domains of the type considered here. 

\section{Discussion}\label{dsec}
We  numerically studied patterns in the 3D Brusselator over boxes with 
Neumann BCs, specifically aiming at snaking branches of steady fronts between patterns, which can also be seen 
as approximations of localized patterns. The basic idea is as in 1D and 2D, 
namely to look for bifurcations from subcritical branches of patterns, 
or from mixed mode branches. However, the numerical challenges are 
significant. 
In 3D, pattern forming systems allow a much larger variety of steady patterns 
than in 1D or 2D. The problem is already quite complicated near onset, 
but on ``nice domains'' (e.g., small cuboids with Neumann BCs) the main branches can be found from (simplified and reduced) amplitude equations. 
Farther from onset, 
there typically is a multitude of patterns, in particular if the domain 
is not very small, and this makes (numerical) continuation and bifurcation 
analysis (much) harder than in 1D or 2D, essentially due to uncontrolled 
branch jumping in the continuation. 

Therefore we focused on the simplest 
situations of small domains in the form of long but slender rods, with an 
underlying BCC lattice, and thus on specific localized patterns, namely 
localized BCCs, and fronts between BCCs and tubes (or localized BCCs 
embedded in a background of tubes or vice versa). Over larger domains we expect 
a huge variety of additional localized patterns, similar to but still extending 
the 2D examples in, e.g., \cite{uwsnak14,w18}. 
However, even for the minimal domains used, the search for localized patterns 
via continuation and bifurcation (as we did for the localized BCCs in Figs.~\ref{s052} and \ref{s04}) is rather delicate. Thus, to obtain starting 
points for the continuation of BCC-to-tubes fronts 
we found it more robust and efficient to use DNS, with 
the Maxwell point of the amplitude system as a guide for promising parameter 
regimes. 
Finally, we gave one example of a moving front between lamellas and tubes. 
It should be interesting to see whether any such localized patterns can be 
realized experimentally, as, e.g., the 3D Turing patterns presented in 
\cite{EV11}.

%\bibliography{/hh/hubib}
%\bibliography{./hubib}
%\bibliographystyle{plain}
\bibliographystyle{alpha}
\input{bru3Dr1.bbl}

\end{document}

%% file: hudefmin.tex
\def\pa{{\partial}}\def\lam{\lambda}\def\noi{\noindent}
\def\del{\delta}
\def\CO{{\cal O}}\def\al{\alpha}\def\ga{\gamma}\def\kap{\kappa}
\newcommand{\barr}{\begin{array}}\newcommand{\earr}{\end{array}}
\newcommand{\bpm}{\begin{pmatrix}}\newcommand{\epm}{\end{pmatrix}}
\def\dd{\, {\rm d}}\def\ri{{\rm i}}

\def\res{{\rm Res}}\def\er{{\rm e}}

\def\Om{\Omega}

\newcommand{\R}{{\mathbb R}}\newcommand{\C}{{\mathbb C}}
\newcommand{\N}{{\mathbb N}}

\newcommand{\reff}[1]{(\ref{#1})}\newcommand{\ov}[1]{{\overline {#1}}}
\def\brem{\begin{remark}}\def\erem{\end{remark}}
\newcommand{\bce}{\begin{center}}\newcommand{\ece}{\end{center}}
\newcommand{\bci}{\begin{compactitem}}\newcommand{\eci}{\end{compactitem}}
\newcommand{\bcen}{\begin{compactenum}}\newcommand{\ecen}{\end{compactenum}}
\newcommand{\hs}[1]{{\hspace{#1}}}\newcommand{\vs}[1]{{\vspace{#1}}}

\def\eps{\varepsilon}

%% file: bru3Dr1.bbl
\newcommand{\etalchar}[1]{$^{#1}$}